\documentclass[xcolor=pdftex,dvipsnames,usenames,prd,nofootinbib,12point]{revtex4}

\input{texdefinitions}

 \newcommand{\ud}{\mathrm{d}}

\newcommand{\ben}{\begin{displaymath}}
\newcommand{\een}{\end{displaymath}}

\newcommand{\eq}[1]{Eq.~(\ref{#1})}

\newcommand{\bfq}{{\bf q}}\newcommand{\bfz}{{\bf z}}
\newcommand{\bfk}{{\bf k}}                                                            
\newcommand{\bfb}{{\bf b}}\newcommand{\bfn}{{\bf n}}

\newcommand {\boldgamma}{\mbox{\boldmath$\gamma$}}

\newcommand{\half}{{\textstyle\frac{1}{2}}}
\newcommand{\svec}[1]{\mbox{\boldmath{$\scriptstyle #1$}}}
\newcommand{\tvec}[1]{\mbox{\boldmath{$#1$}}}
\begin{document}

\title{Electron  Structure: Shape, Size and GPDs in QED\\\hskip7cm NT@UW-14-20}
%\vspace{0.5em}}
\author{Gerald A. Miller}
\address{University of Washington, Seattle WA 9819-1560}
\date{\today}
\begin{abstract} The shape of the electron is studied using lowest-order perturbation theory.  Quantities used to  probe  the structure of the proton: 
form factors, generalized parton distributions, transverse densities,  Wigner distributions and the angular momentum content are computed for the electron-photon component of the electron wave function. The influence of longitudinally polarized photons, demanded by the need for infrared regularization via a non-zero photon mass, is included. The appropriate value of the photon mass depends on experimental conditions, and consequently the size of the electron (as defined by the slope of its Dirac form factor)  bound in a hydrogen atom is  found to be about four times larger than when the electron is free. The shape of the electron, as determined from the transverse  density  and generalized parton distributions is shown to  not be round, and  the free electron is shown to be far less round   than the bound  electron. An electron distribution function (analogous to the 
quark distribution function) is defined, and that of the bound electron is shown to be suppressed compared to that of the free electron.
If the relative transverse momentum of the virtual electron and photon is large compared with the electron mass, the virtual electron and photon each carry nearly the total angular momentum of the physical electron (1/2), with the orbital angular momentum being nearly (-1/2). Including  the non-zero photon mass leads to the suppression of end-point contributions to form factors. Implications for proton structure and color transparency are discussed.

\end{abstract}
\maketitle
\vspace{-3.0em}
%\tableofcontents
%\newpage

\section{Introduction}

Recent experimental progress in measuring improved upper limits for the electric dipole moment of the electron 
\cite{Baron:2013eja,Hinds:2014} has focussed  attention on the non-spherical shape of the electron. These new limits have been interpreted in terms of the deviation of the shape of the electron from spherical symmetry.
A non-zero dipole moment arises from forces that violate parity and time reversal invariance, and its measurement at a much larger value than the standard model prediction would represent new physics.
However, a  non-spherical shape of the electron arises naturally within the standard model via the standard effects of quantum electrodynamics (QED). In particular, the electron fluctuates into a virtual photon and virtual electron, giving the electron both a non-zero spatial extent and a non-spherical shape. These density fluctuations are of first-order in the fine structure constant, $\a$. A schematic representation of the wave function of the electron is given by
\bea |e\rangle=|e_B\rangle +\epsilon |e_B\gamma\rangle,\eea
where $\epsilon^2\propto \a$. We shall refer to the state $|e\rangle$ as the physical electron and the electron in the  $|e_B\gamma\rangle$ 
component as a virtual electron.

The primary purpose of this paper is to investigate the structure of the electron within QED. Quantities such as the extent of the electron and its orbital  angular momentum can only be described within field theory if the light-front or infinite momentum frame approach is used~\cite{Kogut:1969xa,Brodsky:1997de}, and we will use that approach here.
The effects of pair-production from the vacuum are eliminated in this approach, allowing one to  use Fock-space expansions of wave functions.
In the light front approach the longitudinal and transverse  momenta  of the constituents are treated differently, with the longitudinal expressed in  terms of the ratio $x$ with $0\le x\le1$ of the constituent plus-component of the momentum to the total plus component of the momentum, and the transverse momentum $\bfk$ allowed all real values. Thus one refers to roundness as azimuthal symmetry or dependence only on $\bfk^2$.  The use of the spin degree of freedom is needed to perceive a lack of roundness, see for example~\cite{Miller:2003sa}.

The light-front  formalism has been extensively and successfully applied to understand the structure of the proton, and  a variety of  formal  tools have been developed.
These tools include transverse charge distributions\cite{Miller:2007uy,Miller:2010nz,Carlson:2007xd}, generalized parton distributions
\cite{Ji:1998pc}-\cite{Meissner:2009ww}
transverse momentum distributions
\cite{Mulders:1995dh}-\cite{Avakian:2009jt},
and 
Wigner distributions~\cite{Balazs:1983hk}-\cite{Lorce:2011kd}. It is our intention to use these very same tools to elucidate the structure of the electron. This approach has already been used 
in~\cite{Hoyer:2009sg} to provide a qualitative explanation of the sign of the anomalous magnetic moment of the electron and also in~\cite{Burkardt:2008ua} to study the angular momentum content.

A secondary purpose of this paper is use the advantages of knowing the wave function to provide insights useful  for constructing models of the nucleon wave function. The $|e_B\g\rangle$ component of the electron is  analogous to a quark-vector-diquark component of the nucleon. Refs.~\cite{Brodsky:2003pw,Cloet:2012cy} used this analogy to model nucleon electromagnetic form factors.

The  lowest-order, two-component formulation of the electron  has been used for a long time~\cite{Brodsky:1980zm,Brodsky:2000ii}, but one detail is added here. It is standard procedure to let the virtual photon  have a mass (here, $\mu$) to handle the infrared regularization, with  the specific value of this mass depending  on experimental conditions~\cite{Yennie:1961ad,Peskin:1995ev}, or
  the atomic conditions for a bound state~\cite{Eides:2000xc}.
   In particular,  for an electron bound in an atom $\mu\approx 18 \a^2m$, with $m$ the election mass~\cite{Bethe:1947id}
and $\a^2m$ is the Rydberg constant. For a free electron, the value of $\mu$ is determined by the energy resolution~\cite{Yennie:1961ad,Peskin:1995ev}, for which we take $\mu=m$ as a typical value.
It is worthwhile to explain how it is that the photon mass ends up being replaced by the energy resolution. Peskin \& Schroeder~\cite{Peskin:1995ev}  show, for on-shell (free) electrons, that  
$F_1 $ is infrared  (IR) divergent unless  the virtual photon has a mass and the resulting elastic electron-target  cross section contains this regulated  IR divergence.  
But the measured cross section is the sum of the elastic and brehmstrahlung cross sections, and the IR divergence is removed when the sum is taken. The summed cross section is IR safe but depends on the experimental resolution. The net result is that the correct answer is is obtained by
using the computed value of $F_1$ with the photon mass replaced with the energy resolution. Modern experimental facilities have excellent energy resolution (see for example~\cite{Dudek:2012vr})
 so we use the value of the electron mass, 0.51 MeV, which is very small, as the energy resolution. We shall see that  for any values of $\mu>0.002 m$, that there is very little variation with $\mu$, so taking $\mu=m$ is a reasonable simplification.

The Lamb shift has been interpreted as arising from an increase in the spatial extent of the bound electron relative to the free one~\cite{Eides:2000xc}.  This means that there could be substantial differences between the structure of a bound electron with $\mu=18\a^2m$ and a free one with $\mu=m$. The difference in structure between bound and free nucleons is the cause of the EMC effect~\cite{Aubert:1983xm,Hen:2013oha}. Thus  the electron, with its  known wave function  has interesting properties that can lend insight into nucleon properties, and 
 exploring the dependence of the various  electron structure metrics on $\mu$ is an essential part of this paper.  
 
 The remainder of the paper follows the outline: Sect.~II is concerned with a brief review of the light-front tools discussed above. Then our treatment of the electron wave function is described in Sect.~III.  Sect.IV  is concerned with the Dirac form factor, and its dependence on the value of the mass of the virtual photon. Sect. V is concerned with the Pauli form factor. The transverse charge distributions  and end point effects are discussed in in Sect.~VI.  The  generalized parton distribution  $\tilde{F}$ is shown to vanish in Sect.~VII.  Wigner distributions are discussed in Sect.~VIII. The angular momentum content is discussed in Sect.~IX. A summary and discussion Sect.~X concludes the paper.

 \section{Generalized Parton Distributions, Transverse Momentum Distributions and the Wigner Distribution}
 
 Our aim in this Section  is to briefly review the various distributions used to describe the structure of the nucleon,  restricting  the discussion to  the case 
  of skewness 0 (momentum transfer only in the transverse direction). These distributions  are used to describe the electron in later Sections.
 
The generalized parton distributions of twist
two~\cite{Ji:1998pc,Diehl:2001pm,Diehl:2003ny,Ji:2004gf,Diehl:2005jf}  are given by
\begin{eqnarray}
  \label{gpd-def}
F(x,t) &=& \int \frac{dz^-}{8\pi}\, e^{i x p^+ z^-/2}\,
    \langle p'|\, \bar{\psi}(-\half z) \gamma^+ \psi(\half z) |\,
    p\rangle \,\Big|_{z^+ =0,\, \svec{z}=0} 
 =\frac{1}{2p^+} [
  H(x,t)\, \bar{u} \gamma^+ u +
  E(x,t)\, \bar{u} \frac{i \sigma^{+\alpha} \Delta_\alpha}{2m} u
  \, ] ,
  \eea
  where $\psi $ represents the electron field operator, and  $p^+={p'}^+$.
The integral of the above quantities over $x$ give the elastic form factors $F_{1}$ for $E$ and $F_2$  for $H$. There are also~\cite{Diehl:2005jf} 
 \bea
\tilde{F}(x,t) &=& \int \frac{dz^-}{8\pi}\, e^{i x p^+ z^-/2}\,
    \langle p'|\, \bar{\psi}(-\half z) \gamma^+ \gamma_5\, \psi(\half z) |\,
    p\rangle \,\Big|_{z^+ =0,\, \svec{z}=0}  
=\frac{1}{2p^+} [
  \tilde{H}(x,t)\, \bar{u} \gamma^+ \gamma_5 u  
  \, ] ,
\label{tft} \\[0.2em]
F^{j}_T(x,t) &=& -i \int \frac{dz^-}{8\pi}\, e^{i x p^+ z^-/2}\,
    \langle p'|\, \bar{\psi}(-\half z)\, \sigma^{+j} \gamma_5\,
    \psi(\half z) |\, p\rangle \,\Big|_{z^+ =0,\, \svec{z}=0} 
\nonumber \\[0.2em]
 &=& -\frac{i}{2p^+}  [
  H_T(x,t)\, \bar{u} \sigma^{+j} \gamma_5\, u 
  + \tilde{H}_T(x,t)\, \bar{u} \frac{\epsilon^{+j\alpha\beta}
    \Delta_\alpha P_\beta}{m^2} u
+ E_T(x,t)\, \bar{u} \frac{\epsilon^{+j\alpha\beta}
    \Delta_\alpha \gamma_\beta}{2m} u 
  \, ] .
\label{fj}\end{eqnarray}
The integral of $\tilde{F}(x)$ gives the axial form factor, when weighted with the appropriate charges.
Our notation for a four-vector $V$  is $V^{\pm} = (V^0 \pm V^3)$,
  with   transverse part:
$\tvec{V} = (V^1, V^2)$.  Scalar products of boldface vectors obey $\tvec{V}^2 \ge 0$. 
 Roman indices $i$, $j$, $k$ denote 
 the two transverse directions.   The kinematical variables are   $\Delta\equiv p'-p$, $t \equiv \Delta^2$, with $p^+={p'}^+$. The  
 electron  mass  is denoted by $m$.   The polarization of the electron states $\langle p'|$
and $|p\rangle$ are not labelled   here and  the momentum and polarization labels
of the initial and final electron  spinors $\bar{u}$ and $u$ are omitted. We  work only to  first order in $\a$, so no 
Wilson line between the electron  field and its conjugate is needed to preserve gauge invariance.

Transforming the generalized parton distributions at  skewness 0 to impact parameter space yields functions that have a probability
interpretation~\cite{Burkardt:2000za,Miller:2007uy,Miller:2010nz}.  A two-dimensional Fourier
transform gives
\begin{eqnarray}
  \label{impact-mat-elements}
F(x,\tvec{b}) &=&
  \int \frac{d^2 \tvec{\Delta}}{(2\pi)^2}\, 
     e^{- i\svec{b} \cdot\svec{\Delta}}\, F(x, -\tvec{\Delta}^2) .
 \end{eqnarray}
  The two-dimensional Fourier transform in~\eq{impact-mat-elements} causes
electron  field  operators to be  evaluated at a definite transverse position~\cite{Burkardt:2000za,Diehl:2001pm}.
  Integration of \eq{impact-mat-elements} over $x$
gives the transverse charge density, which is the matrix element of the quark charge density 
 in a state of definite transverse position.
 The Fourier transforms
$\tilde{F}(x,\tvec{b})$ and $F_T^j(x,\tvec{b})$  are also defined by  a two-dimensional Fourier transform  as in
\eq{impact-mat-elements}.

 Consider the spin states. 
 It is useful to use states of definite
light-cone helicity~\cite{Soper:1972xc} because the light-cone helicity of
a state is invariant under boosts along the $z$-axis.  It is now conventional procedure to  denote the state %use a state with transverse polarization in the $x$-direction:
\begin{equation}
|p,x\rangle=
{1\over\sqrt{2}}( |p,+\rangle +
 |p,-\rangle ),
\label{xpol}
\end{equation}
  as  a state of transverse  ($x$)
 polarization of the electron. 
A state with both longitudinal and transverse polarization is written as
\bea 
|\Lambda,{\bf S}\rangle=\cos({1\over2}\theta)|+\rangle+\sin\theta e^{i\phi}|-\rangle,\eea
with spin vector ${\bf S}=(\sin\theta\cos\phi,\sin\theta,\sin\phi).$
Thus $\bf S$ and $\Lambda=S_z$ characterizes the transverse and longitudinal   polarizations, $\vec S$ being the general polarization vector.

 Impact parameter generalized parton distribution functions are coordinate space densities for transverse coordinates.. One can access transverse momentum space densities
 through transverse momentum distributions~\cite{Mulders:1995dh}  defined by the correlation function
 \bea
 \Phi^\Gamma_{\a\b}(x,\bfk)=\int {dz^-\over 8\pi}{d^2\bfz\over (2\pi)^2}e^{ixp^+z^-/2}e^{-i \bfk\cdot\bfz}\langle p|\bar{\psi}_\b(-1/2z)\G\psi_\a(1/2z)|p\rangle |_{z^+=0},
\eea
where  the Wilson line is again absent, and the proton states  have no transverse momentum. The superscript $\Gamma$ stands for a twist-two Dirac operator $\Gamma=\gamma^+,\gamma^+\gamma_5,i\sigma^{j+}\gamma_5$ with $j=1,2$.  The functions $\Phi_{\a\b}^\Gamma$ are momentum space densities.

Both generalized parton distributions and transverse momentum distributions are  captured by using 
 five-dimensional Wigner distributions (two position and three momentum coordinates)~\cite{Lorce:2011kd} as seen from the infinite momentum frame (IMF)
 ~\cite{Miller:2007uy,Miller:2010nz,Carlson:2007xd},~\cite{Burkardt:2000za,Burkardt:2002hr,Burkardt:2005td,Diehl:2005jf}.

 The Wigner operators for electrons  at a fixed light-cone time $y^+=0$ \cite{Lorce:2011zta} are defined as follows  
\begin{equation}\label{wigner-operator}
\widehat W^{[\Gamma]}(\bfb,\bfk,x)\equiv\frac{1}{4}\int\frac{\ud z^-\,\ud^2\bfz}{(2\pi)^3}\,e^{i(xp^+z^--\bfk\cdot\bfz)}\,\overline{\psi}(y-\tfrac{z}{2})\Gamma \,\psi(y+\tfrac{z}{2})\big|_{z^+=0}
\end{equation}
with $y^\mu=[0,0,\bfb]$, $p^+$ the average nucleon longitudinal momentum and $x=k^+/p^+$ the average fraction of nucleon longitudinal momentum carried by the active quark.  In general, a Wilson line   is needed for  gauge invariance, but we work only  to order  $\a$, so   no Wilson line is needed for the present applications.

Take matrix elements in nucleon  states that differ by a transverse momentum 
to obtain
\begin{equation}
\begin{split}
W^{[\Gamma]}&(\bfq,\bfk,x,\vec S)=\langle p^+,\tfrac{\bfq}{2},\vec S|\widehat W^{[\Gamma]}({\bf 0},\bfk,x)|p^+,-\tfrac{\bfq}{2},\vec S\rangle\\
&=\frac{1}{4}\int\frac{\ud z^-\,\ud^2\bfz}{(2\pi)^3}\,e^{i(xp^+z^-/2-\bfk\cdot\bfz)}\,\langle p^+,\tfrac{\bfq}{2},\vec S|\overline{\psi}(-\tfrac{z}{2})\Gamma \,\psi(\tfrac{z}{2})|p^+,-\tfrac{\bfq}{2},\vec S\rangle\big|_{z^+=0}.
\end{split}
\end{equation}
Then take a two-dimensional Fourier transform so that
\begin{equation}
\rho^{[\Gamma]}(\bfb,\bfk,x,\vec S)=\int\frac{\ud^2\bfq}{(2\pi)^2}\,e^{-i\bfq\cdot\bfb}\,W^{[\Gamma]}(\bfq,\bfk,x,\vec S).\label{rhoW}
\end{equation}
The functions $\rho^{[\Gamma]}(\bfb,\bfk,x,\vec S)$ are the 
generalized  parton densities mentioned above.
Integration of  Wigner distributions over position and/or momentum leads to  probability distributions. For example,
integrating over $\bfb$  simply  sets $\bfq={\bf 0}$,  so the Wigner distributions reduce to the standard TMDs $\Phi^{[\Gamma]}$~\cite{Meissner:2009ww,Lorce:2011dv}
\begin{equation}
\begin{split}
\int\ud^2b\,\rho^{[\Gamma]}(\bfb,\bfk,x,\vec S)&=W^{[\Gamma]}({\bf  0},\bfk,x,\vec S)
=\Phi^{[\Gamma]}(\bfk,x,\vec S),
\end{split}
\end{equation}
which can be interpreted as quark densities in three-dimensional momentum space.
  Integrating over $\bfk$  sets  ${\bf z}$ to ${\bf 0}$,  so the Wigner distributions reduce to two-dimensional Fourier transforms of the standard GPD correlation functions \cite{Meissner:2009ww,Lorce:2011dv}
\begin{equation}
\int\ud^2k\,\rho^{[\Gamma]}(\bfb,\bfk,x,\vec S)=\int\frac{\ud^2\bfq}{(2\pi)^2}\,e^{-i\bfq\cdot\bfb}\,F^{[\Gamma]}(\bfq,x,\vec S)
,\end{equation}
with
\begin{equation}
F^{[\Gamma]}(\bfq,x,\vec S)\equiv\frac{1}{2}\int\frac{\ud z^-}{2\pi}\,e^{ixp^+z^-/2}\,\langle p^+,\tfrac{\bfq}{2},\vec S|\overline{\psi}(-\tfrac{z}{2})\Gamma \,\psi(\tfrac{z}{2})|p^+,-\tfrac{\bfq}{2},\vec S\rangle\big|_{z^+={\bf z}=0}.
\end{equation}
We see that one obtains  quark densities in transverse position and longitudinal momentum space.
  
\section{Electron  Light Cone Wave Function}

The lowest-order light cone wave function of the electron has been known for a long time~\cite{Brodsky:1980zm,Brodsky:2000ii}. Our only addition to the formalism is
to explicitly include the  longitudinal states of the virtual photon, which are necessary to handle the infrared divergences. 
The  light front model of the $|e_B\gamma\rangle$ component of the  electron consists of the following  light front wave function (LFWF)
 \begin{align}
\Phi^{\L}_{\l_e\,\l}(k,p) 
  = \bar{u}(k,\l_e)\,\ve^*_\nu(q,\l)\g^\nu \varphi   u_N(p,\L),
\label{eq:nucleon_LFWF1}
\end{align}
where $k,~q,~p$ are the virtual  electron, photon  and electron momentum respectively, where $p=q+k$,
and $\l_e,~\l,~\L$ are the related light front helicities, and spinors $u$ are Lepage-Brodsky spinors~\cite{Lepage:1980fj}.
In our evaluations the mass of the physical electron   and that of the bare electron  are both taken as $m$ because 
we  work to first order in $\a$.
The photon is allowed a mass, $\mu$, to regulate the infrared divergence. The transverse polarization vectors in light cone gauge~\cite{Yan:1973qg} have the form 
\begin{align}
\ve^\nu(q,\l=\pm) = \lf(\ve^+,\ve^-,\vect{\ve}^\l\rg) 
              = \lf(0,~\frac{2\,\vect{\ve}^{\l}\cdot \vect{q}}{q^+},~\vect{\ve}^{\l}\rg),
\qquad \text{where} \qquad
\vect{\ve}^{+} = -\frac{1}{\sqrt{2}}\lf(1,i\rg), \quad \text{and} \quad
\vect{\ve}^{-} =  \frac{1}{\sqrt{2}}\lf(1,-i\rg).
\end{align}
These polarization vectors   satisfy the Lorentz condition $\ve \cdot q =0$.  However, the one for longitudinal massive photons does not satisfy this
condition~\cite{Yan:1973qg}. In the interest of clarity 
we are explicit:
\begin{align}
\ve^\nu(q,+)   &= \frac{1}{\sqrt{2}}\lf(0,\,-2\,\frac{q^1+i\,q^2}{q^+},\, -1,\,     -i\rg), &
\ve^{*\nu}(q,+) &= \frac{1}{\sqrt{2}}\lf(0,\,-2\,\frac{q^1-i\,q^2}{q^+},\, -1,\,\ph{-}i\rg), \\
\ve^\nu(q,-)   &= \frac{1}{\sqrt{2}}\lf(0,\,\ph{-}2\,\frac{q^1-i\,q^2}{q^+},\,\ph{-}1,\,-i\rg), &
\ve^{*\nu}(q,-) &= \frac{1}{\sqrt{2}}\lf(0,\,\ph{-}2\,\frac{q^1+i\,q^2}{q^+},\,\ph{-}1,\,\ph{-}i\rg), \\
\ve^\nu(q,0)   &= \lf(0,\,-2\,\frac{\mu}{q^+},\, 0,\, 0\rg), &
\ve^{*\nu}(q,0) &= \lf(0,\,-2\,\frac{\mu}{q^+},\, 0,\, 0\rg),
\end{align}
where we have also included the longitudinal polarization vectors that are  absent in previous treatments.
In the Lepage-Brodsky~\cite{Lepage:1980fj} light cone convection the dot product is given by
$
a\cdot b = 
 \frac{1}{2}\lf(a^+ b^- + a^- b^+\rg) - \vect{a} \cdot\vect{b}
,$
  so that 
\begin{align}
\ve^*_\nu(q,\l)\g^\nu = \frac{1}{2}\,\ve^{*-}(q,\l)\g^+ - \ve^{*1}(q,\l)\g^1 - \ve^{*2}(q,\l)\g^2.
\end{align}

The light-front helicity components of the LFWFs are defined relative to 
Eq.~\eqref{eq:nucleon_LFWF1} via
\begin{align}
\psi^{\L}_{\l_e\,\l}(x,\vect{k})  = \frac{1}{\sqrt{x(1-x)}}\,\Phi^{\L}_{\l_e\,\l}(k,p).
\end{align}
We find 
\begin{align}
\sqrt{(1-x)}\,\psi^{\ua}_{++}(x,\vect{k}) &= \frac{\sqrt{2}\lf(k^1 - i\,k^2\rg)}{x(1-x)}\varphi, & 
\sqrt{(1-x)}\,\psi^{\ua}_{--}(x,\vect{k}) &= 0, \\[2.0ex]
\sqrt{(1-x)}\,\psi^{\ua}_{+-}(x,\vect{k}) &= -\frac{\sqrt{2}\lf(k^1 + i\,k^2\rg)}{1-x}\varphi, & 
\sqrt{(1-x)}\,\psi^{\ua}_{-+}(x,\vect{k}) &= -\sqrt{2}\lf(m - \frac{m}{x}\rg)\varphi , \\[2.0ex]
\sqrt{(1-x)}\,\psi^{\ua}_{+0}(x,\vect{k}) &= -\frac{2\mu}{1-x}\,\varphi, & 
\sqrt{(1-x)}\,\psi^{\ua}_{-0}(x,\vect{k}) &= 0,\label{waveup}
\end{align}
\begin{align}
\sqrt{(1-x)}\,\psi^{\da}_{++}(x,\vect{k} ) &= 0, & 
\sqrt{(1-x)}\,\psi^{\da}_{--}(x,\vect{k} ) &= -\frac{\sqrt{2}\lf(k^1 + i\,k^2\rg)}{x(1-x)}\varphi, \\[2.0ex]
\sqrt{(1-x)}\,\psi^{\da}_{+-}(x,\vect{k} ) &= -\sqrt{2}\lf(m - \frac{m}{x}\rg)\varphi , & 
\sqrt{(1-x)}\,\psi^{\da}_{-+}(x,\vect{k} ) &=  \frac{\sqrt{2}\lf(k^1 - i\,k^2\rg)}{1-x}\varphi, \\[2.0ex]
\sqrt{(1-x)}\,\psi^{\da}_{+0}(x,\vect{k}    ) &= 0, & 
\sqrt{(1-x)}\,\psi^{\da}_{-0}(x,\vect{k} ) &= -\frac{2\mu}{1-x}\,\varphi,
\end{align}
 The $\varphi$ and $\varphi'$ wavefunctions are given by
\begin{align}
\varphi  = \frac{e\sqrt{1-x}}{\lf(m_0^2 - m^2\rg) }, \qquad \text{and} \qquad
\varphi' = \frac{e\sqrt{1-x}}{\lf(m_0'^2 -m^2\rg) },
\end{align}
with
\begin{align}
m_0^2  = \frac{(\bfk-{1\over2}(1-x)\bfq)^2 + m^2}{x}
      + \frac{(\bfk - \frac{1}{2}(1-x)\bfq)^2 + \mu^2}{1-x}, \no \\[1.0ex]
m_0'^2 = \frac{(\bfk + \frac{1}{2}(1-x)\bfq)^2 + m^2}{x}
      + \frac{ (\bfk + \frac{1}{2}(1-x)\bfq)^2 + \mu^2}{1-x},
\end{align}
We have taken the initial electron to have momentum $-\bfq/2$ in writing the above equations.
With the exception of the terms $\psi^\Lambda_{\l_e,0}$,
these wave functions have been known for a long time~\cite{Brodsky:1980zm,Brodsky:2000ii}.

For states with transverse spin in the $x$ direction  (as in \eq{xpol}) we  use the wave function~\cite{Hoyer:2009sg}
 \bea \Psi_{\l_e,\l}\equiv{1\over\sqrt{2}}(\psi_{\l_e,\l}^\ua +\psi_{\l_e,\l}^\da)\eea
These are
\bea\label{bigpsi}
\sqrt{(1-x)}\,\Psi_{++}(x,\vect{k} ) &= \frac{  \lf(k^1 - i\,k^2\rg)}{x(1-x)}\varphi,\no\\
\sqrt{(1-x)}\,\Psi_{+-}(x,\vect{k}) &= -\frac{ \lf(k^1 + i\,k^2\rg)}{1-x}\varphi  +\lf(m{(1 - {x})\over x}\rg)\varphi\no\\
\sqrt{(1-x)}\,\Psi_{+0}(x,\vect{k}) &= -\frac{\sqrt{2}\mu}{1-x}\,\varphi, \no\\
\sqrt{(1-x)}\,\Psi_{-+}(x,\vect{k}) &=  \frac{ \lf(k^1 - i\,k^2\rg)}{1-x}\varphi+\lf(m{(1-{x}\over x}\rg)\varphi , \no\\
\sqrt{(1-x)}\,\Psi_{--}(x,\vect{k}) &= -\frac{\lf(k^1 + i\,k^2\rg)}{x(1-x)}\varphi,\no\\
\sqrt{(1-x)}\,\Psi_{-0}(x,\vect{k}) &= -\frac{\sqrt{2}\mu}{1-x}\,\varphi,
\eea

 \newcommand{\bfkp}{{\bf k}+{1\over2}(1-x){\bf q}}
\newcommand{\bfkm}{{\bf k}-{1\over2}(1-x){\bf q}}
\newcommand{\Karg}{mb\sqrt{1+{\mu^2\over m^2}{x\over (1-x)^2}}}

 \section{$F_1$ of the Electron With An Infrared Regulator}
 
 Our first step is to verify our light cone wave function by computing $F_1$ and comparing with text-book results~\cite{Peskin:1995ev}.
 The form factor $F_1$ is defined as  
 \bea F_1(Q^2)={1\over 2p^+}\langle p,\ua\vert J^+\vert p,\ua\rangle,\eea
 with $J^\mu$ as the electromagnetic current operator, and $Q^2=-q^2$ where $q^\mu$ is the four-momentum transferred to the electron.
 Renormalize using~\cite{Peskin:1995ev}
 \bea F_1(Q^2)=1+ \delta F_1(Q^2)-\delta F_1(0),
  \eea where $\d F_1$ is the first-order correction to $F_1$.
  Then the   renormalized form factor $F_1(Q^2)$ is given by
 \bea F_1(Q^2)=1+{e^2\over 16\pi^3}\int_0^1dx\int d^2k\sum_{\lambda_q,\lambda}[\psi^{\ua*}_{\lambda_q,\lambda}(x,\bfkp)\psi^\ua_{\lambda_q,\lambda}(x,\bfkm)-|\psi_{\lambda_q,\lambda}^\ua(x,\bfk)|^2].\eea
Using our wave functions yields:
 \bea&& F_1(Q^2)=1+{8\pi ^2\alpha\over 16\pi^3}\int_0^1dx  x^2(1-x)I(x,Q^2)\label{f1}\\
 &&I(x,Q^2)={1+x^2\over x^2(1-x)^2}\int {d^2k\over\pi}\left[(k^2-(1-x)^2{Q^2\over4})\chi_f\chi_i-k^2\chi^2\right]+\left(m^2{(1-x)^2\over x^2}+2{\mu^2\over (1-x)^2}\right)\int {d^2k\over \pi}(\chi_f\chi_i-\chi^2),\no\\
 \eea
 where the subscripts $i,(f)$ refer to the initial (final) transverse momentum of the struck virtual electron, and
 \bea\chi_f \equiv {\vp'\over x(1-x)}, \,\chi_i\equiv{\vp \over x(1-x)},\,\chi\equiv{1\over k^2+m^2(1-x)^2+\mu^2x}.\eea
Then
\bea&& I(x,Q^2)=\left(m^2{(1-x)^2\over x^2}+2{\mu^2\over (1-x)^2}\right)I_1(x,Q^2)+{1+x^2\over x^2(1-x)^2}I_2(x,Q^2)-{1+x^2\over x^2}{Q^2\over4}I_3(x,Q^2),\label{I}\\
&&I_1(x,Q^2)\equiv\int d^2k(\chi_f\chi_i-\chi^2),\\
&&I_2(x,Q^2)\equiv\int d^2kk^2(\chi_f\chi_i-\chi^2),\\
&&I_3(x,Q^2)\equiv\int d^2k\chi_f\chi_i.
\eea
The integrals $I_{1,2,3}$ are performed by combining the denominators using the Feynman trick,integrating over $d^2k$ and the Feynman parameter. The results  are 
\bea
&&I_1(x,Q^2)={2\over \sqrt{AB}}{\rm Tanh}^{-1}({\sqrt{AB}\over 2A})-{1\over {\cal M}^2},\label{i1}\\
&& {\cal M}^2\equiv m^2(1-x)^2+\mu^2x,\,A\equiv  {\cal M}^2+(1-x)^2{Q^2\over4},\,B\equiv (1-x)^2Q^2,\\
&&I_2(x,Q^2)=1-2\sqrt{A\over B}{\rm Tanh}^{-1}({\sqrt{AB}\over 2A}),\\
&&I_3(x,Q^2)={2\over \sqrt{AB}}{\rm Tanh}^{-1}({\sqrt{AB}\over 2A}).\label{i3}
\eea
Now that $I(x,Q^2)$ is known via \eq{I}, the expression \eq{f1} for $F_1(Q^2)$ can be evaluated with a one-dimensional  numerical integral.
This expression differs from previous ones via the inclusion of the effects of the longitudinal photon through the terms proportional to $\mu^2$.

To check that this expression is correct we compare with the one from Peskin \& Schroeder. This is their eq(6.56),
\bea && F_1(Q^2)=1+{\a\over2\pi}\int_0^1dx\,dy\,dz\delta(x+y+z-1)\no\\
&&\times\left[\log {m^2(1-z)^2+\mu^2 z\over m^2(1-z)^2+\mu^2z+Q^2x\,y}+{m^2(1-4z+z^2)-Q^2(1-x)(1-y)\over m^2(1-z)^2+Q^2x\,y+\mu^2z}-{m^2(1-4z+z^2)\over m^2(1-z)^2+\mu^2z}\right],\label{PS}
\eea
 amended to retain the terms   $\mu^2 z$ in the $\log$ term of \eq{PS} as stated in the text. 
 
\begin{figure}[h]
\includegraphics[width=10.5cm,height=9.5cm]{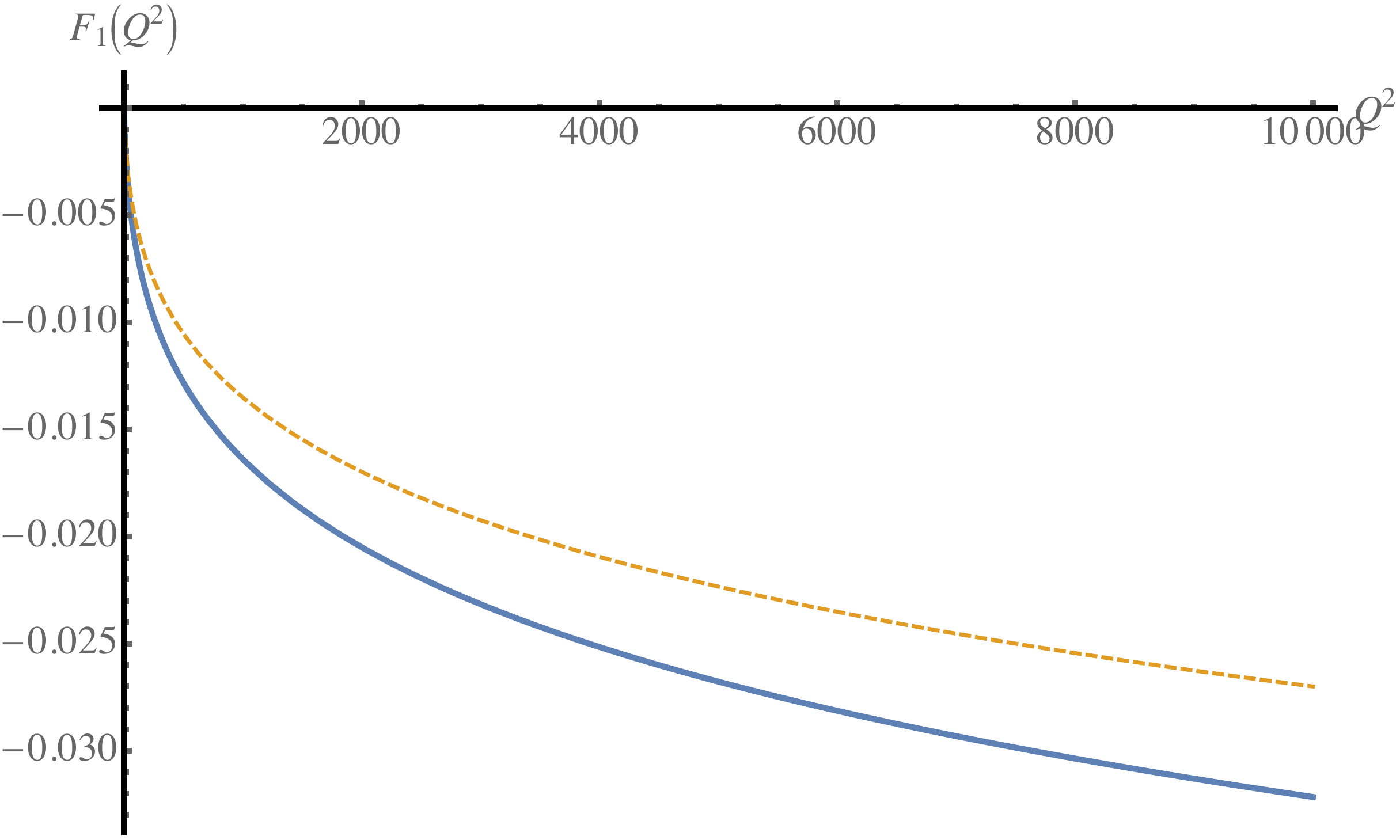}
\caption{(color online)$1-F(Q^2)$ vs. $Q^2$ for $\mu^2=m^2$. Solid (blue) curve: full calculation. Dashed (orange)  ignores longitudinally polarized virtual photons.  $Q^2$ is displayed in terms of dimensionless units $(Q^2/m^2$ is written as $Q^2$ in the figure.}\label{f1fig}\end{figure}

 Using  either  \eq{f1} or \eq{PS} gives numerically identical 
results shown in Fig.~\ref{f1fig}. We see the asymptotic $\log^2Q^2$ dependence, and also the substantial effect of longitudinally polarized virtual photons. 
For $Q^2\gg m^2$ the dependence on $\mu^2$ is essentially $\log Q^2/\mu^2$, and potentially large effects of changing the value of $\mu^2$ are not well-displayed in Fig.~\ref{f1fig}.

For small values of $Q^2$ such that  $Q^2/m^2\ll1,\, \,F_1(Q^2)$ is written as 
\bea F_1(Q^2)\approx 1-\langle b^2\rangle {Q^2\over 4},\eea
where (as we shall see in the next Sect.) $\langle b^2\rangle =\int d^2b\, b^2\rho(b)$. To find $ \langle b^2\rangle $  use  \eq{i1}-\eq{i3} to obtain% an approximation valid for very small values of $Q^2/m^2$
%  \bea F_1(Q^2)\approx
% 1-Q^2{\a\over2\pi}\int_0^1dx\left[\left(m^2(1-x)^5+2 \mu^2x^2(1-x)\right){m^2\over 6{\cal M}^4}+{m^2\over3}{(1+x^2)(1-x)\over {\cal M}^2}\right]\label{fq}\eea
% or
\newcommand{\bmu}{\bar{\mu}}
%\newcommand{\bfz}{{\bf z}}
%$\bmu$
\bea &\langle b^2\rangle =4{\a\over2\pi}\int_0^1dx\left[\left(m^2(1-x)^5+2 \mu^2x^2(1-x)\right){m^2\over 6{\cal M}^4}+{m^2\over3}{(1+x^2)(1-x)\over {\cal M}^2}\right]\\
 &= {\a\over 3\pi}\frac{2 \sqrt{\bmu ^2} \left(\left(\bmu ^2 \left(50-7 \bmu ^2\right)-94\right) \bmu
   ^2+48\right) \left(\tan ^{-1}\left(\frac{1}{\sqrt{\frac{4-\bmu ^2}{\bmu
   ^2}}}\right)-\tan ^{-1}\left(\frac{\bmu ^2-2}{\sqrt{\bmu ^2} \sqrt{4-\bmu
   ^2}}\right)\right)-\sqrt{4-\bmu ^2} \left(7 \left(2 \bmu ^2-7\right) \bmu ^2-\left(\bmu
   ^2-4\right) \left(\left(7 \bmu ^2-8\right) \bmu ^2+4\right) \log \left(\bmu
   ^2\right)+20\right)}{2 \left(4-\bmu ^2\right)^{3/2}},\nonumber\\&\label{b2}
  \eea
  where $\bmu\equiv \mu/m$.
    
  For small values of $\bmu^2$ we may use the approximation
  \bea
  \langle b^2m^2\rangle \approx {\a\over 3\pi}\left(\left(3-\frac{19 \pi }{4}\right) \left(\bmu ^2\right)^{3/2}+3 \pi  \sqrt{\bmu ^2}+\bmu ^2
   \left(4 \log \left(\bmu ^2\right)-\frac{1}{2}\right)-2 \log \left(\bmu ^2\right)+\bmu ^4
   \left(\frac{71}{8}-\frac{7 \log \left(\bmu ^2\right)}{2}\right)-\frac{5}{2}\right)\eea
   which is accurate for $\bmu^2<0.4$. This shows a dramatic $\log$ dependence.
   
\begin{figure}[h]
\includegraphics[width=9.4991cm,height=9.5cm]{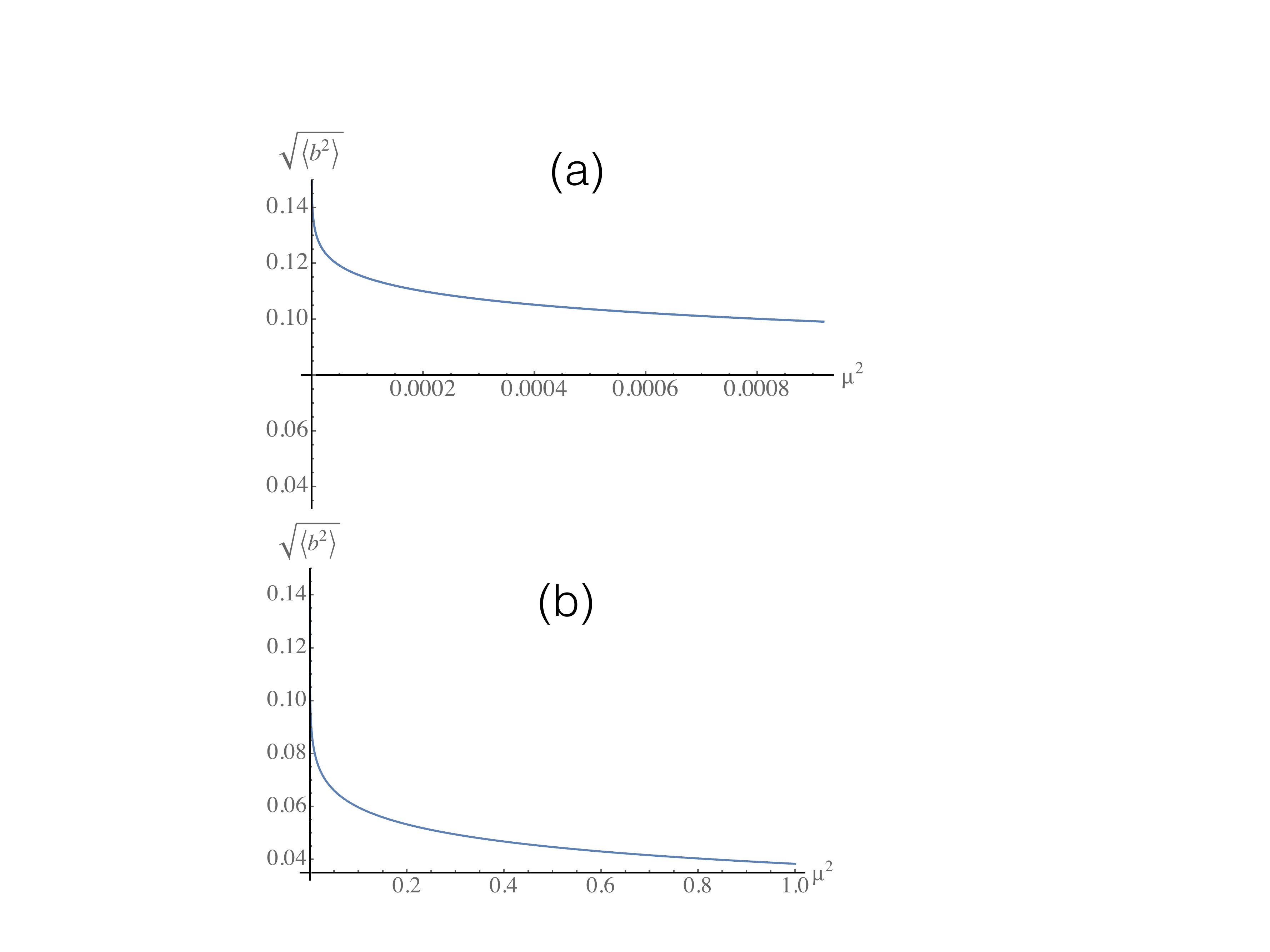}
\caption{(color online) $\mu^2$ dependence of electron radius. Both $\langle b^2\rangle$ and $\mu^2$ are displayed in terms of dimensionless units.}(a) Small values of $\mu^2$,(b)larger values of $\mu^2$\label{bsqmusq}\end{figure}

  A plot of $ \langle b^2\rangle ^{1/2}$ is shown in Fig.~\ref{bsqmusq}. The lower limit on $\mu^2=18\a^2m^2$ corresponds with the appropriate value for electrons bound in hydrogen. The upper limit of $\mu =m$ is taken 
  from typical experimental resolutions on the electron energy in electron scattering experiments. 
  The electron ``size" that enters in interpreting the Lamb shift is defined in terms of the slope of $F_1$ at $Q^2=0$~\cite{Eides:2000xc}. This size, denoted by $R$ is given by $R^2={3\over 2}\langle b^2\rangle.$ Note that the proton size that enters in atomic spectroscopy calculations is 
  determined by the slope of the proton's electromagnetic form factor $G_E$~\cite{Eides:2000xc}.
We see that 
  the size  $\langle b^2\rangle^{1/2}$ of the electron varies from about $0.14/m$ to $0.038/m$ as $\mu^2$ ranges from its lowest to highest values. This is about a factor of four variation. Most of the variation is from the $\log \mu^2$ term.  Remarkably, the bound electron is about four times  larger than the free electron. 
  \section{$F_2$ of the electron with infrared regulator}
 The Pauli form factor $F_2$ is given by 
 \bea&&{F_2(Q^2)\over 2m}(q^1+iq^2)={1\over 2p^+}\langle p',\da\vert J^+\vert p,\ua\rangle,\\
  &&{-F_2(Q^2)\over 2m}(q^1-iq^2)={1\over 2p^+}\langle p',\ua\vert J^+\vert p,\da\rangle,
 \eea
 where $p'=p+q$, $q^+=0,Q^2=\bfq^2,\,q^-=Q^2/M$ and $p$ corresponds to a proton at rest.
 These matrix elements have neither an ultraviolet or infrared divergence, so  {\it if} we take $\mu$ to be zero we
 recover the standard results. However, in the interest of examining implications for nucleon structure, and as preparation for the computation of the transverse charge density of a transversely polarized electron, we present results for non-zero values of $\mu$.
 Using the above wave functions we find
 \bea&F_2(Q^2,\mu^2)={2\a m^2\over\pi}\int dx x\int b\,db J_0(Qb)K_0^2(mb\sqrt{1+{\mu^2\over m^2}{x\over(1-x)^2}}),\eea
 which in taking $\mu$ to zero and integrating by parts reduces to the standard expression:
 \bea F_2(Q^2,\mu^2=0)={4\a m^3\over\pi Q}\int dx x\int b\,db J_1(Qb)K_0(mb)K_1(mb).\eea

\begin{figure}[h]
\includegraphics[width=9.4991cm,height=8.5cm]{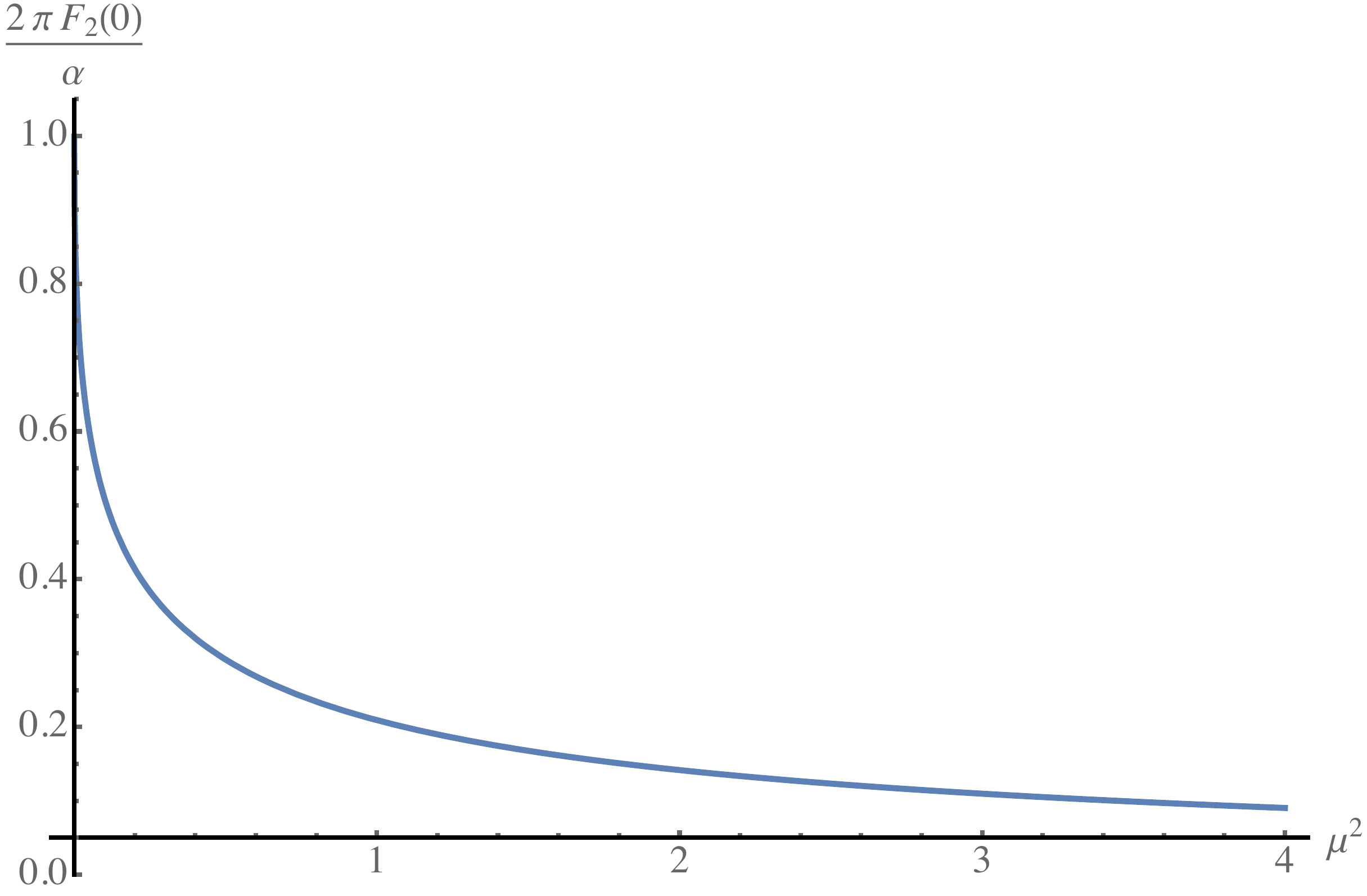}
\caption{(color online) $\mu^2$ dependence of the electron anomalous magnetic moment. Both $\langle b^2\rangle$ and $\mu^2$ are displayed in terms of dimensionless units.}\label{f2}\end{figure}

The very strong $\mu^2$ dependence of $F_2(0)$ is shown in Fig.~\ref{f2}. If $\mu$ represents a di-quark mass (with roughly twice a constituent quark mass)  then using $\mu^2=4m^2$ is appropriate and gives about a factor of five suppression.
For larger values of $Q^2$, the $\mu^2$ dependence is weaker.
 \section{Transverse Charge Density of the Electron}
 The usual procedure is to  obtain the transverse charge density  $\rho(b)$  from the two-dimensional Fourier transform of $F_{1,2}$ or with equivalent manipulations on the appropriate  GPDs. 
 This can  not be done in the present case because of the asymptotic behavior of $F_1(Q^2)\sim \log^2(Q^2)$
 To see this,
 rewrite \eq{f1} as \bea
 F_1(Q^2)=1+{\a\over 2\pi}\int_0^1dxJ(x,Q^2),\\
 J(x,Q^2)\equiv x^2(1-x)I(x,Q^2)\eea 
 The GPD $\rho_0(x,b)$ is the two-dimensional Fourier transform of ${\a\over 2\pi}J(x,Q^2).$ It is well known that asymptotically $J(x,Q^2)\sim \log Q^2,$   arising from the asymptotic form of  $I_2(Q^2)$,
so the Fourier transform into $b$ space does not exist.

 The GPD transverse charge density of an electron polarized in the transverse direction $x$ is given by
\bea &\rho_x(\bfb)=\int {d^2q\over(2\pi)^2}e^{-i\bfq\cdot\bfb}\langle p',x\vert {J^+\over 2p^+}\vert p,x\rangle= \int {d^2q\over(2\pi)^2}e^{-i\bfq\cdot\bfb}(F_1(Q^2)+{iq_y\over m}F_2(Q^2)),
\eea
where $p'=p+q$, $q^+=0,Q^2=\bfq^2,\,q^-=Q^2/M$ and $p$ corresponds to a proton at rest.
We find
\bea \rho_x(\bfb)=\int_0^1\,dx \rho_x(x,\bfb),\eea
with 
\bea &\rho_0(x,b)={\alpha m^2\over 2\pi^2}\left[{1+x^2\over 1-x}(1+{\mu^2\over m^2}{x\over (1-x)^2})K_1^2(mb\sqrt{1+{\mu^2\over m^2}{x\over (1-x)^2}})+((1-x)+2{\mu^2\over m^2}{x^2\over(1-x)^3})K_0^2(mb\sqrt{1+{\mu^2\over m^2}{x\over (1-x)^2}})\right]\\
&\rho_x(x,\bfb)=\rho_0(x,b)+{\alpha m^2\over \pi^2}x \sin\phi_b K_0(mb\sqrt{1+{\mu^2\over m^2}{x\over (1-x)^2}})K_1(mb\sqrt{1+{\mu^2\over m^2}{x\over (1-x)^2}})
\eea
 This   reduces to the result of~\cite{Hoyer:2009sg} if $\mu=0$. 

\begin{figure}[h]
\includegraphics[width=6.5cm,height=6.5cm]{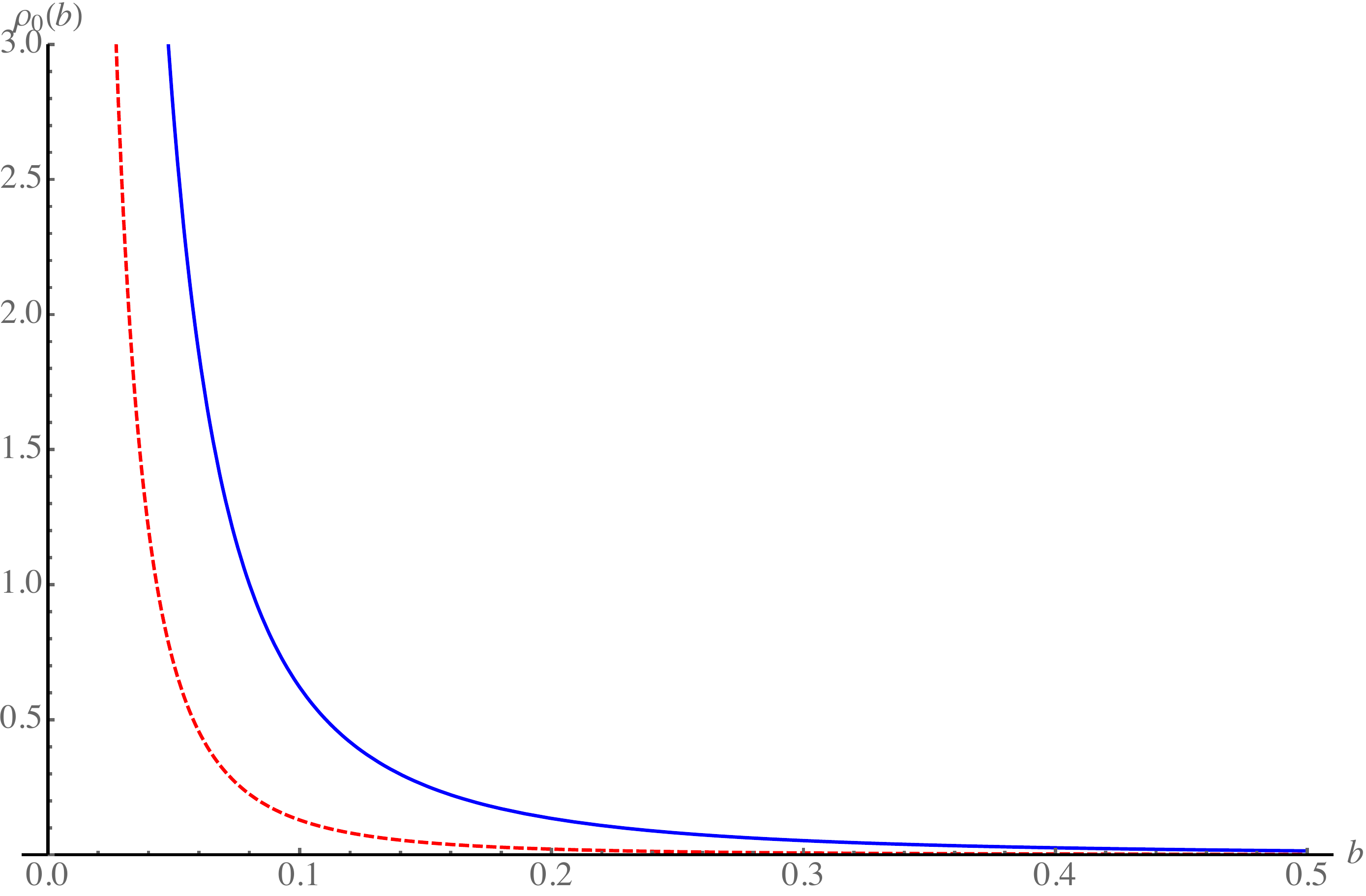}
\caption{(Color online) $\rho_0(b)$ for $\mu=18\a^2$ (solid) bound electron and $\mu=1 $ (dashed)  free electron.    Dimensionless units are used.}\label{rho0}\end{figure}

These densities are well behaved as $x$ approaches unity for non-zero values of $\mu$. 
It is important to understand this if one wants to understand the implications for models of proton structure.
The end-point singularities are severely suppressed  by significant values of $\mu$ which for the proton case correspond to di-quark masses. To see this we note that as $x$ approaches unity, the squares of the modified Bessel functions contain an exponential suppression factor: $e^{-{2\mu b\over 1-x}}$.
Thus, for  non-zero values of $b$,  we may numerically integrate over $x$ to obtain transverse densities. 
Results are shown in Fig~\ref{rho0} for two values of $\mu=18\a^2,1$ corresponding respectively to a bound and free electron.
Observe a strong dependence on the value of $\mu^2$ with the lower value corresponding to an electron of larger spatial extent. This is consistent with the results of the previous section.  The displayed behavior for small values of $b$  is roughly consistent with $\rho_0\sim1/b^2$, which originates in the small argument behavior of $K_1$.

\begin{figure}[h]
\includegraphics[width=8.5cm,height=9.5cm]{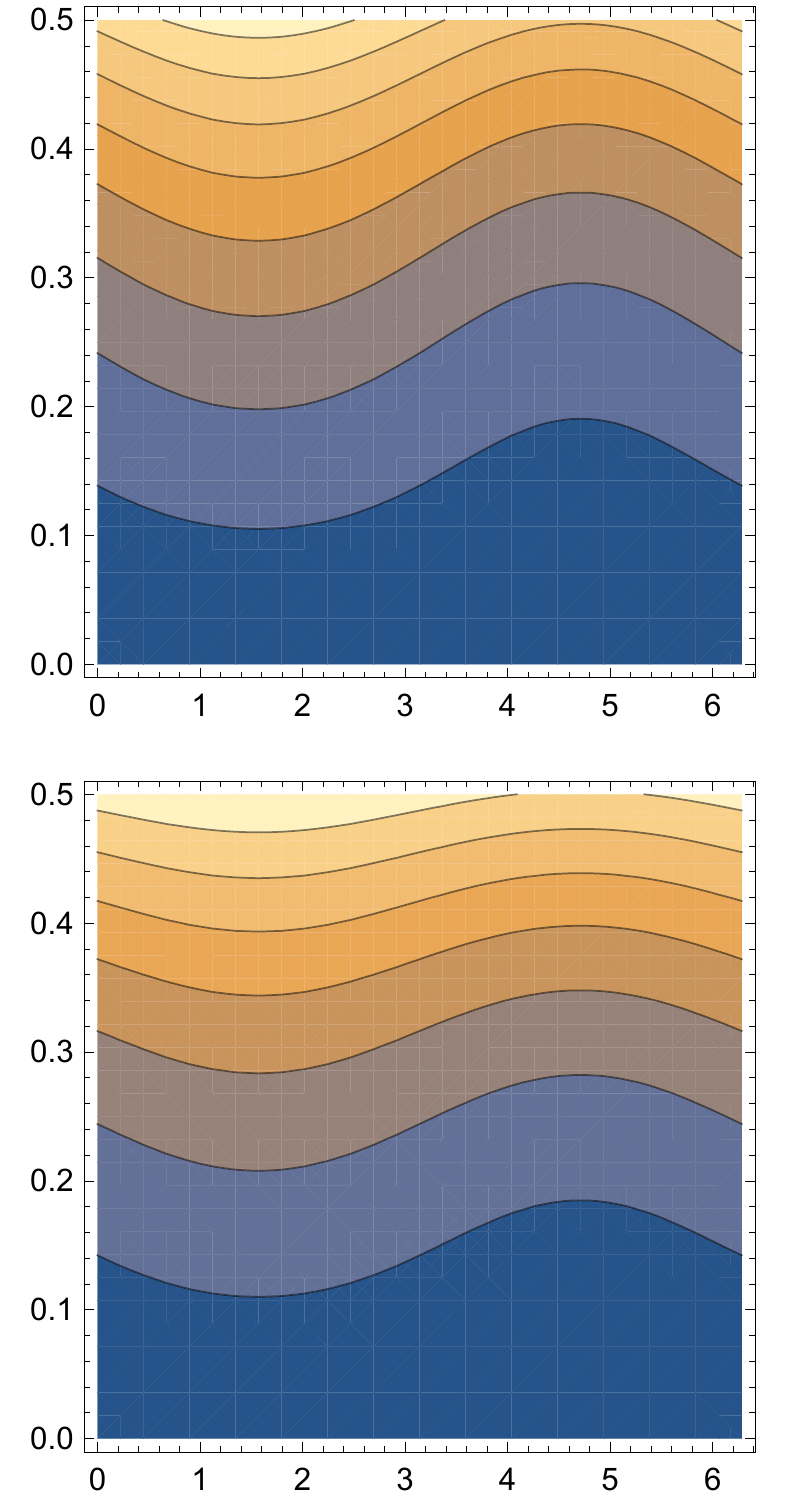}
\caption{(Color online) Dependence of $\rho_x(x,\bfb)$ on $x$ (vertical axis) and $\phi$ horizontal axis for $b=0.1$. Lines of constant $\rho_x(\bfb)$ are  shown.
   Dimensionless units are used. Upper plot  $\mu=18\a^2$, Lower plot $\mu=m$ }\label{Azi}\end{figure}

 Next we wish to examine how much the azimuthal symmetry is violated by $\rho_x(x,\bfb)$. This is shown in 
 Fig.~\ref{Azi}. We see that lines of constant $\rho_x(\bfb)$ have a sinusoidal variation on the polar angle $\phi$.
   This plot shows that the electron charge density is not azimuthally symmetric.  If one uses the spin degree of freedom as a spatial direction, it turns out that the electron is not round.

      \subsection{Use of moments to reconstruct the form factor $F_1$ from $\rho_0(b)$}
  Computing the transverse density at the point at $b=0$ is troublesome because asymptotically $F_1(Q^2)\sim \log^2(Q^2)$,  and so  has no short-distance two-dimensional Fourier transform. Since the usual procedure to obtain the transverse density is to take a two-dimensional Fourier transform, 
one needs another operational procedure to relate the transverse charge density  to $F_1(Q^2)$.
 This is to  take moments
\bea &\langle b^n\rangle\equiv\int_0^1dx \int d^2b b^n\rho_x(\bfb),\,n\ge2.\\
%&={\a m^2\over\pi}\int_0^1 dx\int db b^{n+1} \left[{1+x^2\over 1-x}(1+{\mu^2\over m^2}{x\over (1-x)^2})K_1^2(mb\sqrt{1+{\mu^2\over m^2}{x\over (1-x)^2}})+((1-x)+2{\mu^2\over m^2}{x^2\over(1-x)^3})K_0^2(mb\sqrt{1+{\mu^2\over m^2}{x\over (1-x)^2}})\right]\\
%&={\a \over \pi m^n}\int_0^1 dx{1\over (1+{\mu^2\over m^2}{x\over (1-x)^2})^{{n\over2}+1}}
%\int_0^\infty dy y^{n+1}\left[{1+x^2\over 1-x}(1+{\mu^2\over m^2}{x\over (1-x)^2})K_1^2(y)+((1-x)+2{\mu^2\over m^2}{x^2\over(1-x)^3})K_0^2(y)\right]
\eea
Evaluation leads to 
%\bea%&={\a \over \pi m^n}\int_0^1 dx{1\over (1+{\mu^2\over m^2}{x\over(1-x)^2})^{{n\over2}+1}}\left[{1+x^2\over 1-x}(1+{\mu^2\over m^2}{x\over (1-x)^2})
%\frac{\sqrt{\pi } \Gamma \left(\frac{n}{2}+1\right) \Gamma \left(\frac{n}{2}+2\right) \Gamma
   %\left(\frac{n}{2}\right)}{4 \Gamma \left(\frac{n+3}{2}\right)}+((1-x)+{2\mu^2\over m^2}{x^2\over(1-x)^3})\frac{\sqrt{\pi } \Gamma \left(\frac{n}{2}+1\right)^3}{4 \Gamma \left(\frac{n+3}{2}\right)}\right]\no\\
   %\eea
   %\bea
\bea \langle b^n\rangle   &={\a \over \pi m^n}\frac{\sqrt{\pi } \Gamma \left(\frac{n}{2}+1\right)^3}{4 \Gamma \left(\frac{n+3}{2}\right)}\int_0^1 dx{1\over (1+{\mu^2\over m^2}{x\over(1-x)^2})^{{n\over2}+1}}\left[{1+x^2\over 1-x}(1+{\mu^2\over m^2}{x\over (1-x)^2})
\frac{ n+2}{n}+(1-x+2{\mu^2\over m^2}{x^2\over(1-x)^3})\right]\no\\
\eea
For small $Q^2$, \bea F_1(Q^2)\rightarrow 1-{Q^2\over4}\langle b^2\rangle,\eea
so one needs only $\langle b^2\rangle$, which is given by 
\bea
&\langle  b^2\rangle=%{\a \over \pi m^2}\frac{\sqrt{\pi } \Gamma \left(\frac{2}{2}+1\right)^3}{4 \Gamma \left(\frac{5}{2}\right)}\int_0^1 dx{1\over (1+{\mu^2\over m^2}{x\over(1-x)^2})^{{2\over2}+1}}\left[{1+x^2\over 1-x}(1+{\mu^2\over m^2}{x\over (1-x)^2})
%\frac{ 2+2}{2}+(1-x+2{\mu^2\over m^2}{x^2\over(1-x)^3})\right]\\
%&={\a \over \pi m^2}\frac{\sqrt{\pi } \Gamma \left(2\right)^3}{4 \Gamma \left(\frac{5}{2}\right)}\int_0^1 dx{1\over (1+{\mu^2\over m^2}{x\over(1-x)^2})^{{2\over2}+1}}\left[{1+x^2\over 1-x}(1+{\mu^2\over m^2}{x\over (1-x)^2})
% {2}+(1-x+2{\mu^2\over m^2}{x^2\over(1-x)^3})\right]
% &={\a \over 3 \pi m^2}\int_0^1 dx{1\over (1+ {\mu^2\over m^2}{x\over(1-x)^2})^{2}}\left[{1+x^2\over 1-x}(1+{\mu^2\over m^2}{x\over (1-x)^2})
 %{2}+(1-x+2{\mu^2\over m^2}{x^2\over(1-x)^3})\right]\\
{\a \over \pi m^2}\frac{\sqrt{\pi } \Gamma \left(2\right)^3}{4 \Gamma \left(\frac{5}{2}\right)}\int_0^1 dx{1\over (1+{\mu^2\over m^2}{x\over(1-x)^2})^{{2\over2}+1}}\left[{1+x^2\over 1-x}(1+{\mu^2\over m^2}{x\over (1-x)^2})
 {2}+(1-x+2{\mu^2\over m^2}{x^2\over(1-x)^3})\right]\\
 &={\a \over 3 \pi m^2}\int_0^1 dx\left[{2m^2\over {\cal M}^2}(1+x^2)(1-x)
 +{m^2\over{\cal M}^4}(m^2(1-x)^5+2{\mu^2}{x^2(1-x)}\right].
 \eea
This  value of $\langle b^2\rangle$, obtained as  a moment of $\rho(b)$,
 is the same as obtained from $dF_1(Q^2)/dQ^2$. Indeed,
 analytic integration yields   the result of \eq{b2}.
 %Following result needs to be checked.
 %\bea
%& \langle  b^2\rangle=\frac{1}{3 \sqrt{\pi }}\\
%&\left[
 %-\frac{10 \mu ^4-37 \mu ^2+\left(-5 \mu ^6+26 \mu ^4-28 \mu ^2+16\right) \log \left(\mu ^2\right)+12}{2
  % \left(4-\mu ^2\right)}-\frac{\sqrt{\mu ^2} \left(5 \mu ^6-36 \mu ^4+70 \mu ^2-36\right) \tan
  % ^{-1}\left(\sqrt{\frac{4-\mu ^2}{\mu ^2}}\right)}{\left(4-\mu ^2\right)^{3/2}}\right]
  % \eea
 We reiterate that  
   for electrons bound in an atom $\mu=18 \a^2 m$  but in free space $\mu=m$, as discussed in the Introduction. This leads to the result that the bound electron is about
   4 times larger (in radius) than the free one.

     The utility of the nucleon transverse density is that is directly observable once $F_1(Q^2)$ is sufficiently well known. The purpose of the present subsection is to demonstrate this same observability for the electron transverse density.  
  Usually, integration over $x$ in \eq{gpd-def} leads to
   \bea F_1(Q^2)=\int d^2b J_0(Q b)\rho_0(b).\label{ff1}\eea
 To relate this expression to moments we use the expansion 
   \bea
   J_0(x)=\sum_{n=0}^\infty({x\over2})^{2n}(-1)^n{1\over \Gamma[n+1]^2}.\eea
   The influence of the term with $n=0$ in $F_1$ is removed by the renormalization procedure. Thus use the above sum starting at $n=1$ in  \eq{ff1} to
   find
   \bea F_1(Q^2)=1+ \sum_{n=1}^\infty \left({-Q^2\over4m^2}\right)^n
   {\langle b^{2n}\rangle\over \Gamma[n+1]^2}.\eea
   The alternating series expansion in $Q^2\over 4m^2$ converges rapidly for $Q^2\le 4m^2$, but fails for larger values of $Q^2$. The success of this series in reproducing $F_1(Q^2)$ partially justifies the interpretation of $\rho_0(b)$ as the two-dimensional FT of $F_1(Q^2)$.
   
   \subsection{End point singularities-implications for nucleon structure}
   
The electron-photon component of the physical electron provides an example for building models of nucleon structure  because one component of the nucleon wave function consists of a quark and a vector di-quark. In particular, the transverse size of composite systems is relevant for understanding whether or not color transparency (suppression of initial or final state interactions) in high momentum transfer processes~\cite{Frankfurt:1992dx,Frankfurt:1994hf}. For large values of $Q^2$ at fixed values of $x$, it is expected that only Fock states with transverse size less than about $1/Q$ contribute to 
form factors~\cite{Lepage:1980fj,Frankfurt:1993es}
. 
In this case, interactions with surrounding particles, which depend on the square of the transverse size are suppressed. However, 
for values of near the  the endpoints, $x=0$ or 1, contributions of large transverse size could be relevant. The transverse size of the electron-photon state is the relative distance between the electron and the photon $|\bfb_e-\bfb_\g|=b/(1-x)$ and becomes large for values of $x$ near $1$.

Ref.~\cite{Hoyer:2009sg} addressed this issue, concluding that ``the power suppressed contributions to the electron's Dirac form factor cannot  be restricted to small impact parameters $(b\rightarrow0)$  at any $Q$.".  However, that reference did not include the necessary effect of the virtual photon mass, $\mu$. If one is considering an analogy with the proton wave function the relevant value of $\mu$ corresponds to  roughly twice the mass of the fermion because the  di-quark is  made of 
 two constituent quarks. This mass is not at all small. Moreover, including a non-zero value of $\mu$ enables one to compute quantities that are integrals over all values of $x$.

\begin{figure}[h]
\includegraphics[width=8.5cm,height=9.5cm]{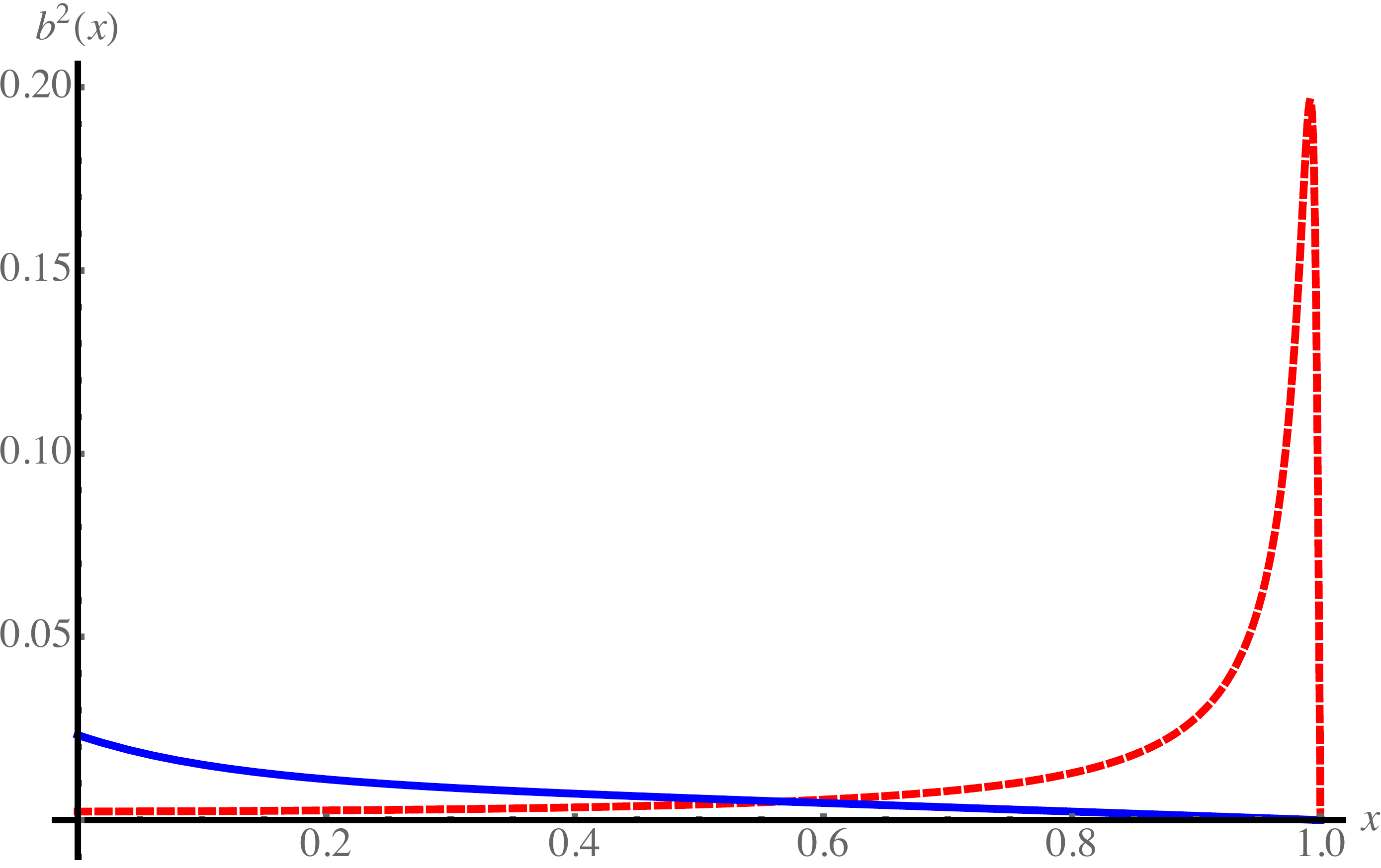}
\caption{(Color online) Dependence of $b^2(x)$ on $x$ for $\mu=2m$ (solid) and $\mu=0.01 m$  (dashed).  Dimensionless units are used. }\label{end}\end{figure}

 This issue is already addressed in Fig.~\ref{rho0}, which shows the $x$-integrated transverse density  for $\mu=18\a^2m$ and 
 $\mu=m$. Both densities are strongly peaked at $b=0$, falling roughly as $1/b^2$ for small values of $b$.
  Increasing the value of $\mu$ makes the electron-photon state more compact. A further increase of $\mu$ to 2$m$ would make the state even more compact. Another way of looking at  
 the importance of end-point contributions is  shown in Fig.~\ref{end} which shows 
\bea b^2(x)\equiv \int d^2b b^2 \rho_x(x,\bfb)\label{b2mu}\eea
  for $\mu=0.01m$ and $\mu=2m$.
 The curve obtained using $\mu=2m$ has been multiplied by a factor of 10 so as to become visible on the plot.
 We see that  there is a significant contribution to $b^2(x)$ at large $x$ if $\mu=18\a^2m$, but this is completely suppressed by using $\mu=2m$. Given Figs~\ref{rho0} and \ref{end} we conclude that  (for Fock state components analogous to the electron-photon component) end point contributions can not be expected to 
give significant contributions to form factors at large values of $Q^2$.

 \section{Shape of the Electron}
 
The  non-spherical shape of the proton   was addressed using spin-dependent 
densities~\cite{Miller:2003sa,Miller:2008sq}
As a spin 1/2 object the proton has no net  quadrupole moment. However, the probability that quark of position $\bfb$ (or momentum $\bfk$ ) and  spin direction $\bfn$   confined in  a proton polarized in a direction ${\bf S}$  does have a wide variety of interesting shapes depending on the directions of the three vectors that are involved. This probability is  the matrix element of a spin-dependent density. Later~\cite{Miller:2007ae}
it was realized that the coordinate space spin-dependent density is related to a generalized parton distribution and the  momentum space distribution is related to a transverse momentum distribution.

Given the known wave function of the electron, we have an opportunity to compute the spin-dependent densities of the electron. 
 We take the electron to be polarized along the $x$ axis, work to order $\alpha$ and compute the matrix element of the relativistic spin-dependent density operator 
\bea {\cal O}^+(\bfn)\equiv {\gamma^+\over 2}\left(1+\bfn\cdot\boldgamma\gamma_5\right)\eea
for a virtual electron  with spin in a direction $\bfn$. % that is transversely polarized in a direction defined to be the $x-$axis.
This is  the electron version of the 
projection operator for quarks of transverse polarization $\bfn$~\cite{Diehl:2005jf}. 

 The related  transverse charge density  $\rho_x(\bfb,\bfn)$ of a physical  electron polarized in the transverse direction $x$ is given by
\bea &\rho_x(\bfb,\bfn)=\int {d^2q\over(2\pi)^2}e^{-i\bfq\cdot\bfb}\langle p',x\vert {{\cal O}^+\over 2p^+}\vert p,x\rangle
%= \int {d^2q\over(2\pi)^2}e^{-i\bfq\cdot\bfb}(F_1(Q^2)+{iq_y\over m}F_2(Q^2)),
={1\over2}\rho_x(\bfb)+{1\over2}\rho_{Tx}(\bfb,\bfn).
\eea
The first term of ${\cal O}^+$ is just the transverse charge density of the previous section. The second term of
${\cal O}^+$ depends on $\bfn$ and gives $\rho_{Tx}(\bfb,\bfn)$.
Evaluation of the matrix element  leads to 
\bea
&\rho_{Tx}(\bfb,\bfn)={e^2\over 16\pi^3}\int {d^2q\over(2\pi)^2}e^{-i\bfq\cdot\bfb}\int_0^1dx\int d^2k\sum_{ \lambda}[\Psi^{*}_{+,\lambda}(x,\bfkp)\Psi_{-\lambda}(x,\bfkm)(n^1-in^2)\nonumber\\&+\Psi^{*}_{-,\lambda}(x,\bfkp)\Psi_{+\lambda}(x,\bfkm)(n^1+in^2)].
\eea
where $p'=p+q$, $q^+=0,Q^2=\bfq^2,\,q^-=Q^2/M$ and $p$ corresponds to a proton at rest.  
  Then evaluation leads to
 \bea
&\rho_{Tx}(\bfb,\bfn)={\a\over 2\pi^2}\int {d^2q\over(2\pi)^2}e^{-i\bfq\cdot\bfb}\int_0^1{dx\over1-x}
\int d^2k\left( [{2(\bfkp)\cdot(\bfkm)\over x(1-x)^2}+2{\mu^2\over(1-x)^2}]n^1+i{m\over x^2}(n^1q^2-n^2q^1)\right)\varphi'\varphi
\eea
Thus we have contributions to the GPDs $E_T'+2\tilde{H}_T$and $H_T$, \eq{fj}.

\begin{figure}[h]
\includegraphics[width=6.5cm,height=6.5cm]{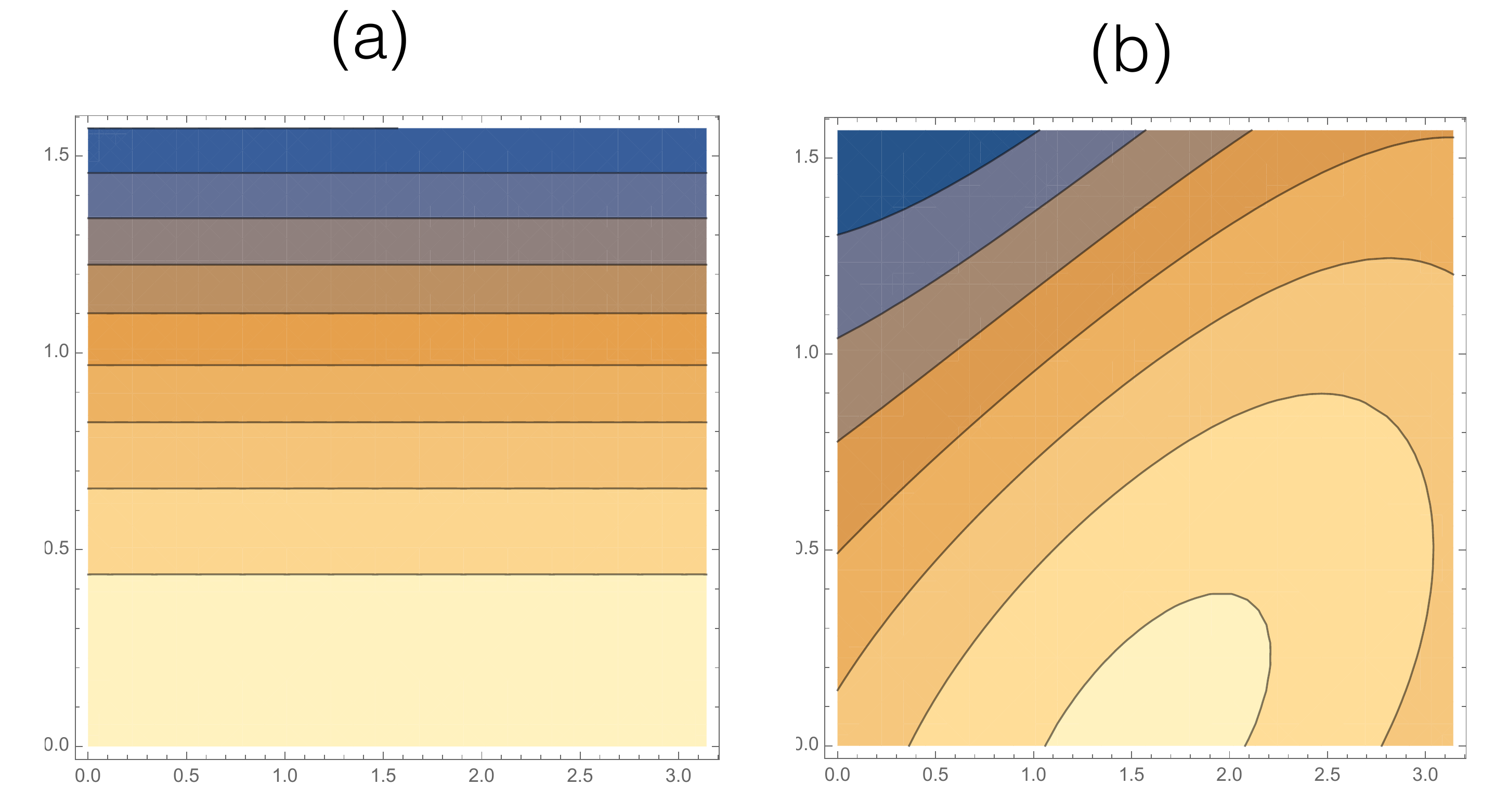}
\caption{(Color online) Dependence of $\rho_{Tx}(x,\bfb)$ on $\phi_n$ (vertical axis) and $\phi$ (horizontal axis) for $b=1,\mu=1$ free electron. Lines of constant $\rho_{Tx}(\bfb)$ are  shown.
   Dimensionless units are used. (a) $\mu=18\a^2$, bound electron (b) $\mu=1$, free electron} \label{rhot}\end{figure}

Evaluating using  the wave functions of \eq{bigpsi} gives
 \bea
&\rho_{Tx}(\bfb,\bfn)={\a\over 2\pi^2}\int_0^1{dx\,x^2\over1-x}\left[ {2m^2(1+{\mu^2\over m^2}{x^2\over(1-x)^2})\over x(1-x)^2}K_1^2(\Karg)+{2\mu^2\over(1-x)^2}K_0^2(\Karg)\right]n^1
\\&+{(n^1b^2-n^2b^1)\over b}{\a\over 2\pi^2}\int_0^1{dx\,x^2\over1-x}{m^2\over x^2}\sqrt{1+{\mu^2\over m^2}{x\over(1-x)^2}}K_0(\Karg)K_1(\Karg),
\eea
The term ${(n^1b^2-n^2b^1)\over b}=n^1\sin\phi-n^2\cos\phi$.
Results are shown in Fig.~\ref{rhot} as lines of constant  $\rho_{Tx}(\bfb,\bfn)$. Examination shows that $\rho_{Tx}$ for a bound electron is roughly independent of $\phi$, but this is not so for the free electron. We conclude that the bound electron is rounder than the free electron!

 \section{Electron GPD $\tilde{F}$}

The only GPD of Sect. II not yet computed is $\tilde{F}$.
  Evaluation leads to 
 \bea
&\tilde{F}(x,\bfb)={\a\over 4\pi^2}\int {d^2q\over(2\pi)^2}e^{-i\bfq\cdot\bfb}\int d^2k\sum_{ \lambda}[\Psi^{*}_{+,\lambda}(x,\bfkp)\Psi_{+\lambda}(x,\bfkm)\nonumber
\\&-\Psi^{*}_{-,\lambda}(x,\bfkp)\Psi_{-\lambda}(x,\bfkm)].
\eea

We find using \eq{bigpsi}
\bea \sum_{ \lambda}\Psi^{*}_{+\lambda}(x,\bfkp)\Psi_{+\lambda}(x,\bfkm)=((1+x^2)\bfk'\cdot\bfk+2\mu^2x^2+m^2(1-x)^4+iq_y\,x(1-x)^3)\chi_f\chi_i\eea
\bea \sum_{ \lambda}\Psi^{*}_{-\lambda}(x,\bfkp)\Psi_{-\lambda}(x,\bfkm)=((1+x^2)\bfk'\cdot\bfk+2\mu^2x^2+m^2(1-x)^4+iq_y\,x(1-x)^3)\chi_f\chi_i\eea
Thus to order $\a$, $\tilde{F}(x,\bfb)=0$.
\section{Wigner Distributions of the electron}

There are many possible Wigner distributions (Sect.~II) for the operators $\Gamma$ and each has its own related transverse densities and TMDs.  We shall only give two examples.

The Wigner distribution ($\g^+$) for a  electron  with positive longitudinal polarization is given by
\bea
W^{[\g^+]}&(\bfq,\bfk,x,\ua)={1\over 16\pi^3}\sum_{\lambda_q,\lambda}[\psi^{\ua*}_{\lambda_q,\lambda}(x,\bfkp)\psi^\ua_{\lambda_q,\lambda}(x,\bfkm) ].\label{We}\eea
The related Fourier transform from $\bfq$ to $\bfb$ leads to the transverse densities treated extensively above.

\subsection{Electron distribution function}

\begin{figure}[h]
\includegraphics[width=6.5cm,height=6.5cm]{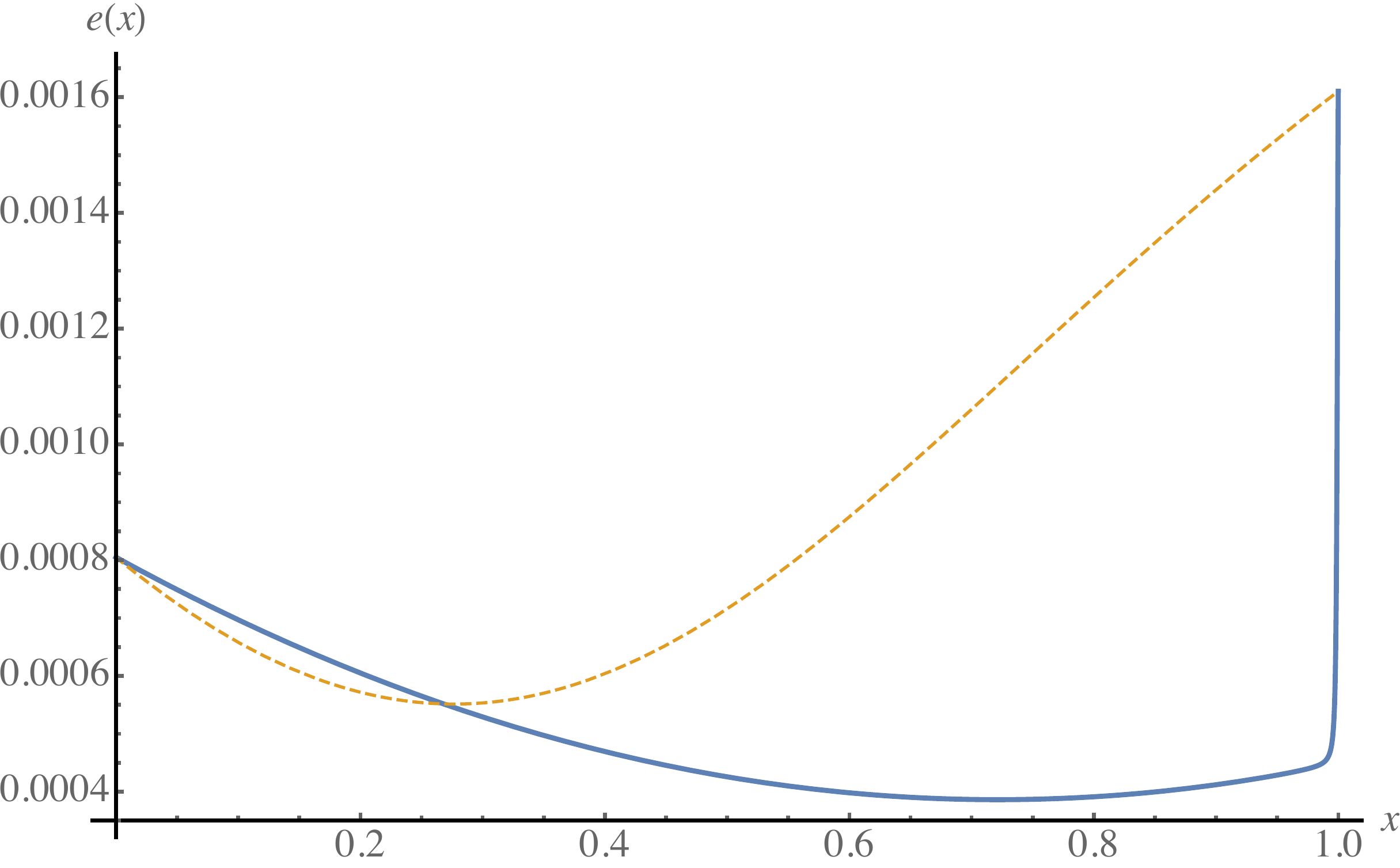}
\caption{(Color online) Electron distribution function $e(x)$ $\rho_0(b)$ for $\mu=18\a^2$ (solid) (bound electron) and $\mu=1 $ (dashed) free electron.}\label{Dist}\end{figure}

To get the transverse momentum distribution TMD $\Phi^{\g^+}(\bfk,x,\ua)$,
set $\bfq$ to 0. Using \eq{We}  yields 
\bea 
\Phi^{\g^+}(\bfk,x,\ua)= {2\alpha\over 4\pi^2}\left[{\bfk^2(1+x^2) }+m^2(1-x)^4+{2\mu^2x^2}\right]{1\over (\bfk^2+{\cal M}^2)^2}
\eea
Further integration over $\bfk$ would  yield the analog of a quark distribution function.  The integral over $\bfk$ diverges and must be renormalized. Following conventional procedure~\cite{Brodsky:2000ii}, we use Pauli-Villars regularization and integrate on $\bfk$ to an upper limit, $\L$, which can be a function of $x$. Thus we define the electron distribution function in analogy to the usual quark distribution function:
\bea e(x,\L)\equiv \int d^2k  \Theta(\L^2-\bfk^2)\,\Phi^{\g^+}(\bfk,x,\ua).\eea
For illustrative purposes we take $\L={\cal M}(x)$. This gives 
\bea 
e(x)\equiv e(x,\L={\cal M})={\a\over 2\pi} [(1+x^2)(\ln2-{1\over2})+{(m^2(1-x)^2+2\mu^2 x^2 \over{\cal M}^2}]
.\eea
The function $e(x)$ is plotted in Fig.~\ref{Dist}. We see a difference between bound and free electrons that is reminiscent of the EMC effect~\cite{Aubert:1983xm}: the distribution function is suppressed for large values of $x$, except for $x=1$. At $x=1$ $e(x=1)={\a\over 2\pi}(2(\ln2-1/2)+1), $ independent of the value of $\mu$.

\subsection{Position and momentum together}
Ref.~\cite{Lorce:2011kd} showed that integration of  $\rho^\Gamma(\bfb,\bfk,x,\vec{S})$ over $b_y$ and $k_x$ gives the probability for the virtual electron to have given values of $b_x,k_y$,
$\rho^\Gamma(b_x,k_y,x,\vec{S})$ . This is possible because $b_x$ and $k_y$ are not canonically conjugate variables and not constrained by  the uncertainty principle.  Use \eq{We} in \eq{rhoW} to  find
\bea&
\rho^{\g^+}(\bfb,\bfk,x,\ua)=\int{d^2\bfq\over (2\pi)^2}e^{-i\bfq\cdot\bfb}{2\alpha\over 4\pi^2}\left[{(\bfkp)\cdot(\bfkm)(1+x^2) }+m^2(1-x)^4+{2\mu^2x^2}\right]\nonumber\\&
\times{1\over ((\bfkp)^2+{\cal M}^2) ((\bfkm)^2+{\cal M}^2)}\eea
%Use
%\bea \chi(k)\equiv{1\over \bfk^2+{\cal M}^2}=\int{d^2\bfk\over(2\pi)}e^{-i\bfk\cdot\bfb}\chi(b),\,\chi(b)=K_0({\cal M}b)
%\eea
Evaluation leads to the result
\bea\rho^{\g^+}(b_x,k_y,x,\ua)&= \Theta(1-x){\a\over 4\pi}{1\over (1-x)}e^{-2|B_x|\sqrt{{\cal M}^2+k_y^2}}[{m^2(1-x)^4+2\mu^2x^2\over k_y^2+{\cal M}^2}+(1+x^2)({k_y^2\over {\cal M}^2+k_y^2})]
\eea
for $x<1$,
where  ${\bf B}\equiv ({\bfb\over1-x})$   and 
\bea {\cal M}^2=(1-x)^2m^2+\mu^2x.\eea

\begin{figure}[h]
\includegraphics[width=8.5cm,height=8.5cm]{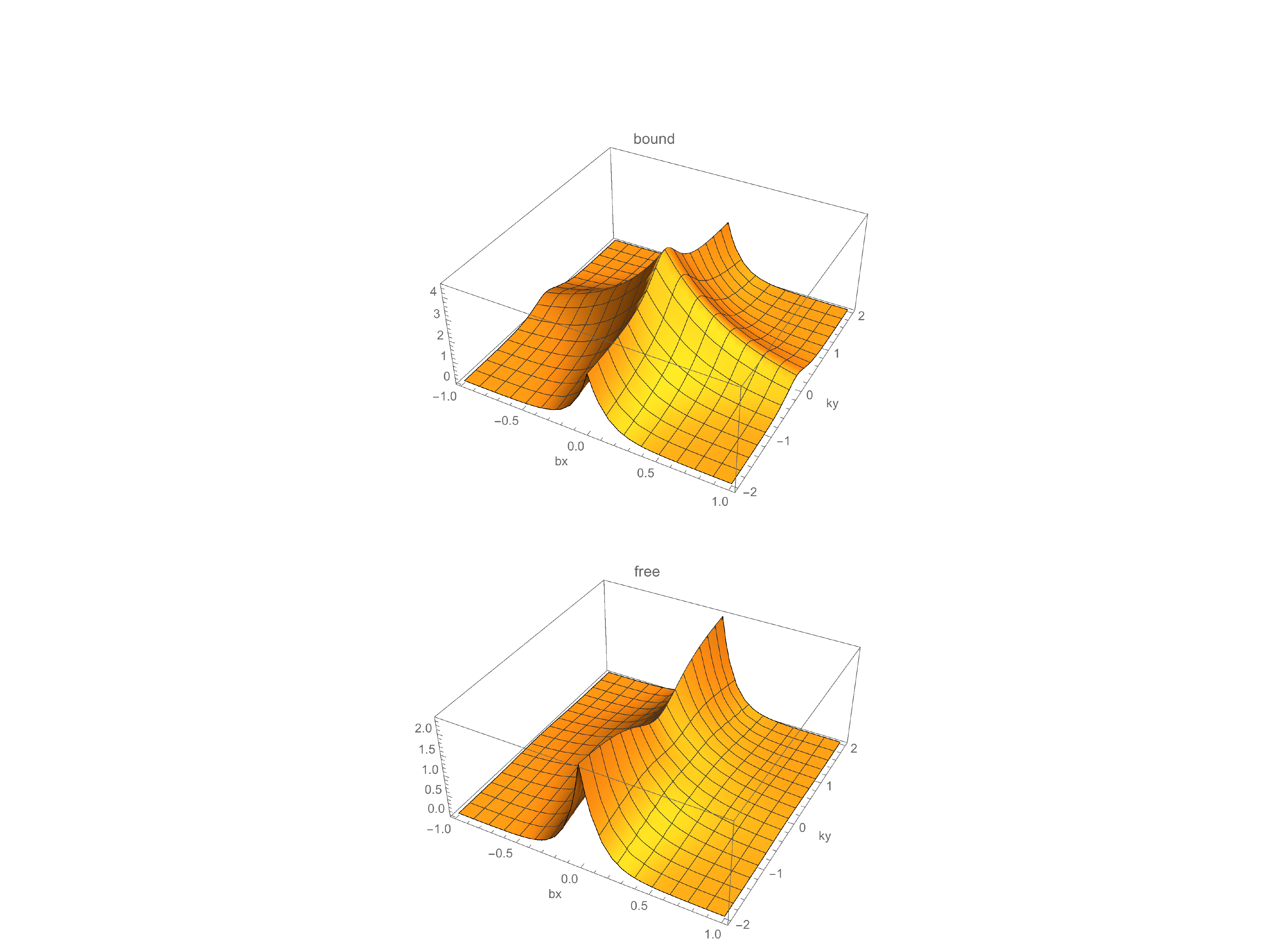}
\caption{(Color online)  $\rho^{\gamma^+}(b_x,k_y, x=0.5,\ua)$. Upper panel: $\mu=18\a m$ (bound electron). Lower panel: $\mu=m$ (free electron). Dimensionless units are used. }\label{wigner}\end{figure}

The function $\rho^{\g^+}(b_x,k_y,x,\ua)$ for $x=0.5$ is displayed in Fig.~\ref{wigner}. 
If one considers a bound electron, the extent in $mb_x$ is comparable to that of $k_y/m$, but for a free electron (lower panel) 
 the extent in $k_y/m$ is much broader than the extent in $mb_x$.   
 If $x=1$  $\rho^{\g^+}(b_x,k_y,x,\ua)$  contains a factor
$\d(b_x)$  reflecting the point-like nature of the bare electron.

The derivation above implies that one may obtain $\rho^{\g^+}(b_y,k_x,x,\ua)$  from $\rho^{\g^+}(b_x,k_y,x,\ua)$ by the interchange $(b_x,k_y)\leftrightarrow(b_y,k_x)$.
 \section{Angular Momentum Content}
 
 We want to  examine the angular momentum content of the electron. For the electron-photon state there are three contributions: the spin of the virtual electron, the spin of the virtual photon and  the orbital angular momentum.  This topic was 
 taken up in Ref.~\cite{Hoyer:2009sg} in terms of the transverse coordinate $\bfb$.   Here we use transverse momentum $\bfk$. Additionally the inclusion of the virtual photon mass $\mu$ allows us to integrate over $x$ and obtain the various contributions as a function of only one variable,  $\bfk^2$.  
  
 The normalization of the ${\cal O}$ Fock state of a spin-up electron  for given values of $x,\bfk$
 is given by
 \bea
& N(\bfk,x)={1\over 16\pi^3}\sum_{{\lambda_e},\lambda}|\psi^\ua_{\lambda_e,\lambda}|^2
\\&= {2\alpha\over 4\pi^2}\left[{\bfk^2(1+x^2) }+m^2(1-x)^4+{2\mu^2x^2}\right]{1\over (\bfk^2+{\cal M}^2)^2}\eea
where the second line is obtained from using \eq{waveup}.
The expectation value of the electron spin $\lambda_e$ in the $|e\gamma\rangle$ Fock state of a parent electron with spin $\ua$ as
\bea&\langle \lambda_e(\bfk,x) \rangle_\ua\equiv {1\over N(\bfk,x)}\sum_{\l_e,\l}\langle \ua;\l_e,\l|S^z_e|\ua;\l_e,\l\rangle\\&
={1\over 2N(\bfk,x)}\sum_{\l}[|\psi^\ua_{+,\l}|^2-|\psi^\ua_{-,\l}|^2],\eea
where $S_z^e={1\over2}\sigma^z$ is the electron spin operator. The expectation value for the photon helicity $\l$ is given by
\bea&\langle \lambda(\bfk,x) \rangle_\ua={1\over N(\bfk,x)}\sum_{\l_e}[|\psi^\ua_{\l_e,+}|^2-|\psi^\ua_{\l_e,-}|^2].\eea
The orbital angular momentum is given by 
\bea&\langle L_z(\bfk,x) \rangle_\ua={1\over N(\bfk,x)}[-|\psi^\ua_{+,+}|^2+|\psi^\ua_{+,-}|^2]\eea

We find:
\bea &N(\bfk,x)\la \lambda_e(\bfk,x)\ra_\ua={1\over2}2{\a\over4\pi^2} \left[{\bfk^2(1+x^2) }-m^2(1-x)^4+{2\mu^2x^2}\right]{1\over (\bfk^2+{\cal M}^2)^2}\\&
N(\bfk,x)\la\l(\bfk,x)\ra_\ua=2{\a\over4\pi^2}[{\bfk^2(1-x^2)}+m^2(1-x)^4]{1\over (\bfk^2+{\cal M}^2)^2}
\\&N(\bfk,x)\la L_z(\bfk,x)\ra_\ua=2{\a\over4\pi^2}{\bfk^2(-1+x^2) }{1\over (\bfk^2+{\cal M}^2)^2}\eea
With these expressions one finds that 
\bea &\la \l_e(\bfk,x)\ra_\ua={1\over2}\frac{\left[{\bfk^2(1+x^2) }-m^2(1-x)^4+{2\mu^2x^2}\right]}{\left[{\bfk^2(1+x^2) }+m^2(1-x)^4+{2\mu^2x^2}\right]}\\&
\la\l(\bfk,x)\ra_\ua=\frac
{[{\bfk^2(1-x^2)}+m^2(1-x)^4]} {\left[{\bfk^2(1+x^2) }+m^2(1-x)^4+{2\mu^2x^2}\right]}
\\&\la L_z(\bfk,x) \ra_\ua=\frac{\bfk^2(-1+x^2)  }{\left[{\bfk^2(1+x^2) }+m^2(1-x)^4+{2\mu^2x^2}\right]}.
\label{s1}\eea
Taking the sum of the three terms of \eq{s1} leads to the correct result that entire total angular momentum of the electron is accounted for:
\bea \la \l_e(\bfk,x)\ra_\ua+\la\l(\bfk,x)\ra_\ua+\la L_z(\bfk,x) \ra_\ua={1\over2}\eea
Note that increasing the photon mass increases the fractional spin carried by the virtual electron, while increasing the electron mass 
decreases the fractional spin carried by the virtual electron, but increases the fractional spin carried by the photon. The orbital angular momentum is negative as found  by~\cite{Hoyer:2009sg}.  

Next we obtain the three contributions as quantities that are integrated over $x$.
Define 
\bea N(\bfk)\equiv \int_0^1\,dx\,N(\bfk,x)\eea and integrate using $\mu^2=1$ to find
\newcommand{\bfK}{{\bf K}} 
\bea N(\bfK)={1\over16\pi^3}\frac{4 \left(\sqrt{4 \bfK^2+3} \bfK^2+\left(-4 \bfK^4+2 \bfK^2+3\right) \cot ^{-1}\left(\sqrt{4
   \bfK^2+3}\right)\right)}{\left(4 \bfK^2+3\right)^{3/2}},\eea
   where $\bfK\equiv\bfk/m$
Then
\bea &N(\bfK)\la \l_e(\bfK)\ra_\ua={1\over2}\frac{4 \left(8 \bfK^6+22 \bfK^4+19 \bfK^2+5\right) \cot ^{-1}\left(\sqrt{4 \bfK^2+3}\right)-2
   \sqrt{4 \bfK^2+3} \left(4 \bfK^4+5 \bfK^2+2\right)}{\left(\bfK^2+1\right) \left(4
   \bfK^2+3\right)^{3/2}}\\
   &N(\bfK)\la\l(\bfK)\ra_\ua=
   {\frac{2 \left(4 \bfK^2+1\right) \left(\sqrt{4 \bfK^2+3}-2 \left(2 \bfK^2+1\right) \cot
   ^{-1}\left(\sqrt{4 \bfK^2+3}\right)\right)}{\left(4 \bfK^2+3\right)^{3/2}}}
\\&N(\bfK)\la l_z(\bfK)\ra_\ua=-\frac{\bfK^2 \left(\left(2 \bfK^2+3\right) \sqrt{4 \bfK^2+3}-8 \bfK^2 \left(\bfK^2+1\right) \cot
   ^{-1}\left(\sqrt{4 \bfK^2+3}\right)\right)}{\left(\bfK^2+1\right) \left(4
   \bfK^2+3\right)^{3/2}}   \eea
   
\begin{figure}[h]
\includegraphics[width=8.4991cm,height=8.5cm]{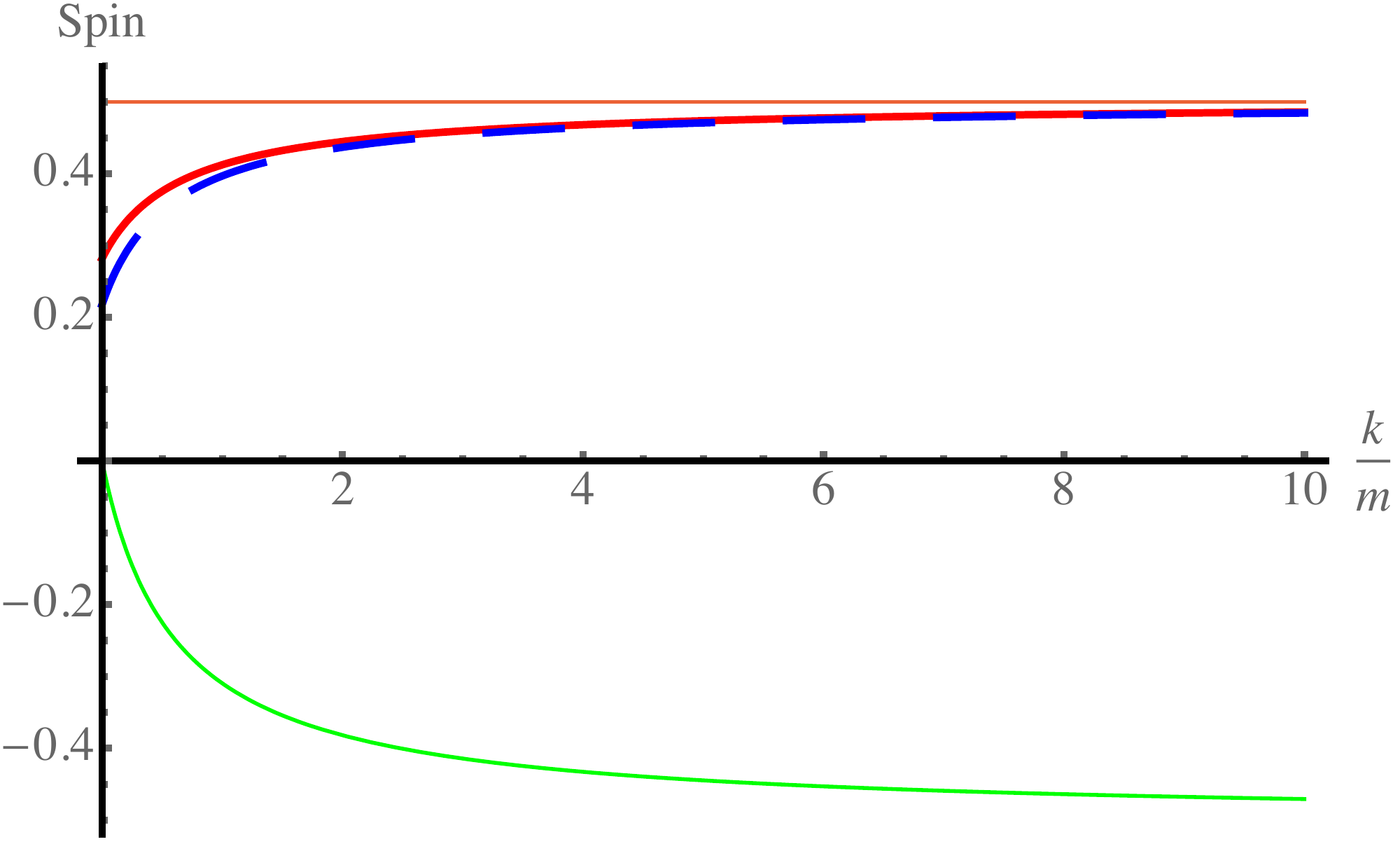}
\caption{(Color online) Spin structure of the electron.}\label{spin}\end{figure}

 Fig.~\ref{spin} shows the numerical results. We see that
   for values $\bfK^2>1$ the photon and virtual electron each  carry almost   the entirety of the spin of the physical electron and the orbital 
   angular momentum is the almost the negative of the spin of the physical electron. 
   %This is from eints.nb
 This feature is also qualitatively obtained for a bound electron with  $\mu^2=L$. 
   
 \section{Summary}
 
This paper is aimed at elucidating the interesting structure of the electron within QED. Lowest-order perturbation theory is used
to obtain the virtual electron-photon component of the physical electron wave function.  The first remark is that indeed this structure is indeed interesting. The application  of tools developed to understand nucleon structure to the electron reveals  several unexpected features. Many quantities related to electron structure can be computed, and we have tried to concentrate on the most interesting ones.

Here we specifically include the effects of longitudinally polarized photons that emerge  through the need to use a non-zero photon mass $\mu$  to treat infrared divergences. This enables us to show that using the light-cone Fock state wave function of the electron reproduces the well-known text book results for the electron Dirac form factor.
(The Pauli form factor has been treated this way for a long  time~\cite{Brodsky:1980zm}.)  

The transverse size of the electron is found to depend strongly on the value of $\mu$, with the bound electron being about four times larger than the free electron. The electron-photon component is found to be very compact, with small values of $b$ dominant in any case. End-point contributions  to form factors are strongly suppressed for values of $\mu$ greater than the fermion mass.
 
 The shape of the electron is addressed through the use of transverse charge densities and generalized parton distributions. The direction of the spin provides an axis so that one can assess the azimuthal symmetry  (in transverse space) of the electron wave function. We find that azimuthal symmetry is not obtained so that the electron is not round. The generalized parton distribution $\tilde F$, \eq{tft},  is shown to vanish, if one uses lowest-order perturbation theory.
 Since the electron wave function is known, one may compute the various Wigner distributions. These depend strongly on the value of $\mu$. In particular, the electron distribution function (analogous to the quark distribution function)  is shown to be strongly suppressed for a bound electron. This is similar to the EMC effect for nucleons bound in nuclei.
 
The angular momentum content of the electron is also interesting. The spin of the electron, the spin of the photon and the relative orbital angular momentum each contribute significantly. For large values of the virtual electron-photon relative momentum, the spin of the electron and spin of the photon add up to almost one unit  angular momentum, with the orbital angular momentum being almost the minus one-half unit  of  angular momentum.

One might ask about the experiments that could directly measure some of the distributions of the electrons that are used here.
These are already included as part of the standard radiative corrections that are normally included in analyzing electron scattering data. 
 \section*{Acknowledgements}
 This material is based upon work supported by the U.S. Department of Energy Office of Science, Office of Basic Energy Sciences program under Award Number DE-FG02-97ER-41014.
 I thank I. Clo\"{e}t for important technical help and S.~D.~Ellis, P.~Hoyer, S.~J.~Brodsky, M.~Eides and M.~Peskin  for useful discussions. 
  %===============================================================================
 %=================================================
% Bibliography
%=================================================
\cleardoublepage

%\footnotesize
%\addcontentsline{toc}{chapter}{\bf Bibliography}

\end{document}